\documentclass[%
 aip,
 amsmath,amssymb,
 reprint,%
]{revtex4-1}

\usepackage{graphicx}
\usepackage{dcolumn}
\usepackage{bm}
\usepackage{makecell} 
\usepackage{booktabs} 

\usepackage[utf8]{inputenc}
\usepackage[T1]{fontenc}
\usepackage{mathptmx}
\usepackage{physics}
\usepackage{siunitx}
\usepackage[version=4]{mhchem}

\usepackage{hyperref}
\hypersetup{
    colorlinks=true,
    linkcolor=blue,
    filecolor=blue,      
    urlcolor=blue,
    citecolor=blue,
}

\usepackage{color,soul}
\usepackage{xspace}
\usepackage{xcolor}
\usepackage{tablefootnote}

\renewcommand{\vec}[1]{\ensuremath{\boldsymbol{\mathbf{#1}}}}

\begin{document}

\preprint{AIP/123-QED}

\title{Phase Transitions Affected by Natural and Forceful Molecular Interconversion}

\author{Thomas J. Longo}
\affiliation{Institute for Physical Science and Technology, University of Maryland, College Park, MD 20742, USA}
\email{tlongo1@umd.edu}

\author{Mikhail A. Anisimov}
\affiliation{Institute for Physical Science and Technology, University of Maryland, College Park, MD 20742, USA}
\affiliation{Department of Chemical and Biomolecular Engineering, University of Maryland, College Park, MD 20742, USA}
\email{anisimov@umd.edu}

\date{\today}

\begin{abstract}
If a binary liquid mixture, composed of two alternative species with equal amounts, is quenched from a high temperature to a low temperature, below the critical point of demixing, then the mixture will phase separate through a process known as spinodal decomposition. However, if the two alternative species are allowed to interconvert, either naturally (\textit{e.g.} the equilibrium interconversion of enantiomers) or forcefully (\textit{e.g.} via an external source of energy or matter), then the process of phase separation may drastically change. In this case, depending on the nature of interconversion, two phenomena could be observed: either phase amplification, the growth of one phase at the expense of another stable phase, or microphase separation, the formation of nongrowing (steady-state) microphase domains. In this work, we phenomenologically generalize the Cahn-Hilliard theory of spinodal decomposition to include molecular interconversion of species and describe the physical properties of systems undergoing either phase amplification or microphase separation. We apply the developed phenomenology to accurately describe the simulation results of three atomistic models that demonstrate phase amplification and/or microphase separation. We also discuss the application of our approach to phase transitions in polyamorphic liquids. Lastly, we describe the effects of fluctuations of the order parameter in the critical region on phase amplification and microphase separation. 
\end{abstract}

\maketitle




\section{Introduction}
Phase transitions play a crucial role in condensed-matter physics and chemistry, astrophysics, biology, and engineering applications \cite{Goldenfeld_Lectures_1992,nicolis_self-organization_1977,Onuki_2002,Mazumdar_Cosmic_2019, Boeynaems_Protein_2018, Hyman_LLBiology_2014,Naumann_Nonlinear_2001}. Examples include structural, superconductive, and ferromagnetic transitions in solids, superfluidity in helium, as well as various liquid-solid, liquid-liquid, and liquid-vapor phase transformations  \cite{,LL_Stat_Phys,anisimov_thermodynamics_2018,puri_kinetics_2004,Halperin_Superconductors_1974,Stanley_PT_1971,Nelson_Solids_1978,Batlogg_Properties_1991,DeGenne_Crystals_1993,Xu_Review_2021}. While equilibrium phase transitions in simple systems are well-studied and understood, the description of phase transformations in the presence of interconversion between alternative molecular or supramolecular states, and in systems far from equilibrium, is much less developed \cite{puri_kinetics_2004,Marro_Nonequilibrium_1999,Muktish_Nonequilibrium_2005}. 

Two distinguished molecular species may separate if the interactions between them are not favorable for mutual mixing. The most recognizable example is almost complete separation of water and oil. Another example is the possible demixing of structural isomers, such as enantiomers with opposite molecular chirality\cite{Lombardo_ThermoMechanism_2009,Latinwo_MolecModel_2016,Petsev_Effect_2021,Uralcan_Interconversion_2020}. In contrast, solid ferromagnetic or ferroelectric materials (in the absence of a magnetic or electric field), have no restriction on the direction of magnetization or polarization, meaning that there is no conservation of the number of magnetic spins or electric dipoles with a particular orientation, and consequently, these materials do not establish equilibrium coexistence between phases with alternative magnetizations or polarizations\cite{hohenberg_theory_1977,Onuki_2002,puri_kinetics_2004}. If fluids exhibit interconversion of species, similar to the flipping of magnetic spins or electric dipoles, the conservation of the number of alternative molecules is broken and the thermodynamics and kinetics of phase separation will dramatically change.

After a binary mixture, initially containing equal amounts of the alternative molecules, is quenched from the one-phase homogeneous region at a high temperature into the unstable region below the critical temperature of demixing, the species will start to separate through a process known as ``spinodal decomposition'' \cite{cahn_phase_1965}. However, if the two species with the same density may rapidly interconvert (like in a mixture of enantiomers\cite{Uralcan_Interconversion_2020}), then to avoid the formation of an energetically unfavorable interface, the phases will compete with each other until one of them is eliminated\cite{Ricci_Computational_2013,Shum_Phase_2021}. In this work, we show that this phenomenon (referred to as ``phase amplification''\cite{Shum_Phase_2021}) is the result of the competition between the diffusive dynamics of phase separation and the ``flipping'' dynamics of interconversion.

In a nonequilibrium system, if an external force causes the alternative molecules to stay in equal numbers (referred to as ``forceful interconversion''\cite{Longo_Structure_2021}), the striking phenomenon of steady-state phase separation into mesoscale domains can be observed. Such nonequilibrium ``microphase separation'' is fundamentally different from the equilibrium mesoscale structures (bicontinuous or modulated) in microemulsions\cite{Gompper_Lattice_1994} or the microphase separation of block and polyelectrolyte copolymers\cite{Jones_Soft_Matter_2002,Borukhov_Polyelectrolyte_2000}, where the mesoscale heterogeneities are the result of the minimization of the equilibrium free energy.\cite{Anisimov_Mesothermo_2010} It is also different in origin from the metastable patterns formed by ``frozen'' spinodal decomposition, such as those observed in glasses\cite{Kim_Metallic_2013,Lin_Nanocrystallization_2015}. Contrarily, the steady-state microphase separation is one of the simplest examples of dissipative structures in condensed matter. In this work, we show that the characteristic length scale of this dissipative structure emerges as a result of the competition between forceful interconversion and mutual diffusion.

We phenomenologically generalize Cahn-Hilliard's theory of spinodal decomposition to include natural (equilibrium) and forceful (nonequilibrium) molecular interconversion. In particular, we quantitatively describe the phenomena of phase amplification and steady-state microphase separation through the analysis of the results obtained by Shumovskyi \textit{et al.}'s hybrid binary-lattice/Ising model\cite{Shum_Phase_2021} and Uralcan \textit{et al.}'s chiral-mixture model\cite{Uralcan_Interconversion_2020}.

The paper is organized as follows. In Sec. \ref{Sec_Theory}, we phenomenologically generalize Cahn-Hilliard's theory to include interconversion of species in equilibrium and dissipative systems. In Sec. \ref{Sec-Sfactor}, we investigate the domain growth for a phase separating system exhibiting molecular interconversion through the time evolution of the structure factor. In Sec. \ref{Sec_Apps}, we compare our approach with the results of simulations of a hybrid binary-lattice/Ising model\cite{Shum_Phase_2021} and a chiral-mixture model \cite{Lombardo_ThermoMechanism_2009,Latinwo_MolecModel_2016,Petsev_Effect_2021,Uralcan_Interconversion_2020}. In Sec. \ref{Sec_Polyamorphism}, we consider a connection between our approach to phase transitions with molecular interconversion to liquid polyamorphism \cite{Stanley_Liquid_2013,Tanaka_Liquid_2020}, a striking phenomenon which could be a result of interconversion between molecular or supramolecular structures \cite{anisimov_thermodynamics_2018,Caupin_Minimal_2021}. In Sec. \ref{Sec_CritFlucts}, we provide scaling arguments to further generalize our approach beyond the mean field approximation to account for the effects of critical fluctuations. In the conclusion, we provide suggestions for the extension of our approach to a broader range of phenomena.

\section{\label{Sec_Theory} Spinodal Decomposition Affected by Natural and Forceful Molecular Interconversion}

In this section we phenomenologically generalize Cahn-Hilliard's theory of spinodal decomposition to include the effects of two types of molecular interconversion: natural, in which the interconversion of species mirrors the equilibrium ``flipping'' of spins in the Ising model, and forceful, in which an external force causes the alternative molecules to stay in equal amounts. This section contains three parts.  In~\ref{Subsec_Thermo_Model}, we provide the thermodynamics basis of our approach. In~\ref{Subsec_KineticEqns}, we discuss the kinetics of phase domain growth. In~\ref{Subsec_PhaseAmpVsMicSep}, we elaborate on the conditions for the observation of phase amplification or microphase separation in simulations and experiment.

\subsection{Chemical Potential for Equilibrium Molecular Interconversion\label{Subsec_Thermo_Model}}


We consider a symmetric binary mixture of two species A and B with concentrations, $c_A$ and $c_B = 1-c_A$. Both species have the same densities ($\rho = 1$), viscosities, and molecular weights. This system can be described by a Landau-Ginzburg free-energy functional with a single order parameter uniquely linked to the concentration of species A as, $\phi = 2(c_A - 1/2)$. This functional reads as
\begin{equation}\label{Eqn_LGfunc}
	F[\{ \phi  \} ] = \frac{1}{\rho} \int_V\left(\hat{G}(\phi,T,P) + \frac{1}{2}\kappa\abs{\grad{\phi}}^2 \right)\dd{V}
\end{equation}
where the first term represents the thermodynamic ``bulk'' free energy and the second term is included to describe the contribution to the free energy due to inhomogeneities within the system. For an isotropic system, the coefficient $\kappa$ is the square of the range of intermolecular interactions, on the order of the square of the molecular size. In the applications to the atomistic models (Sec. \ref{Sec_Apps}), we adopt $\kappa = 1$. 

\begin{figure}[t]
    \centering
    \includegraphics[width=\linewidth]{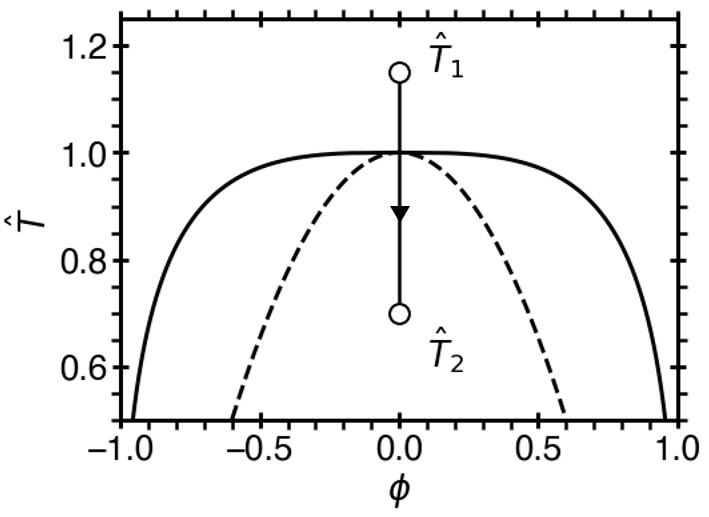}
    \caption{A symmetric binary mixture is quenched along the critical concentration (where the order parameter $\phi=0$) from a point, $\hat{T}_1$ in the one phase region to a point, $\hat{T}_2$ below the critical temperature of demixing in the unstable region. The solid curve indicates the liquid-liquid coexistence (binodal), while the dashed curve indicates the limit of stability (spinodal).}
    \label{Fig_Quench}
\end{figure}

The bulk free energy density for the system, $\hat{G}$, is the reduced Gibbs energy, $\hat{G} = G/k_\text{B}T_\text{c}$, where the critical temperature for liquid-liquid phase separation is $T_\text{c}$, as
\begin{equation}
   \hat{G} = \hat{G}_A + (1-c_A)\hat{G}_{BA} + \hat{G}_\text{mix}
\end{equation}
where $\hat{G}_{BA}=\hat{\mu}_B^0-\hat{\mu}_A^0$ is the reduced difference between the Gibbs energies (chemical potentials) of pure species A and B, referred to as the Gibbs energy change of reaction. For the symmetric binary-lattice (``regular solution'') model, $\hat{G}_\text{mix}$ is formulated through the order parameter, $\phi$, and in the mean field approximation, it reads as
\begin{equation}\label{Eqn_f0}
\begin{split}
    \hat{G}_\text{mix} =& \hat{T}\bigg[\left(\frac{1+\phi}{2}\right)\ln\left(\frac{1+\phi}{2}\right) \\ &+ \left(\frac{1-\phi}{2}\right)\ln\left(\frac{1-\phi}{2}\right)\bigg] + \frac{\epsilon}{4}(1-\phi^2)
\end{split}
\end{equation}
where $\hat{T} = T/T_\text{c}$ and $\epsilon$ is a non-ideality interaction parameter, which generally depends on temperature and pressure. The conditions for liquid-liquid phase equilibrium is 
\begin{equation}
    \pdv{\hat{G}_{\text{mix}}}{\phi} = \frac{\hat{T}}{2}\ln\left(\frac{1+\phi}{1-\phi}\right) -\frac{\epsilon}{2}\phi = 0
\end{equation}
From the stability condition, $\partial^2\hat{G}_\text{mix}/\partial \phi^2 = 0$, the interaction parameter is $\epsilon=2$ in units of $k_BT_\text{c}$, since $T_\text{c} = \epsilon/2k_B$.

We consider the interconversion between molecular states A and B through a reversible chemical reaction of the form
\begin{equation}\label{Eqn-INT}
	A \ce{<=>[$k_1$][$k_2$]} B
\end{equation}
where $k_1$ and $k_2$ are the forward and reverse reaction rates respectively. Since the order parameter could be considered as the reaction coordinate, the chemical reaction equilibrium condition reads
\begin{equation}
    \pdv{\hat{G}}{\phi}\bigg|_{T,P} = \pdv{\hat{G}_\text{mix}}{\phi}\bigg|_{T,P} -\frac{1}{2}\hat{G}_{BA}= 0
\end{equation}
This reaction-equilibrium condition constrains the number of thermodynamic degrees of freedom for the system. Consequently, the fraction of interconversion, given through the order parameter $\phi$, is no longer an independent thermodynamic variable, but instead, becomes a function of temperature and pressure. The reduced Gibbs energy change of reaction, $\hat{G}_{BA}$ can be expressed through the reaction equilibrium constant $\mathcal{K}(T,P) = k_1/k_2$ and given in the linear approximation as
\begin{equation}\label{Eq_GBA}
    \hat{G}_{BA} = -\hat{T}\ln{\mathcal{K}} = \hat{e} + \hat{\upsilon}P-s\hat{T}
\end{equation}
where $\hat{e}=e/k_\text{B}T_\text{c}$ is the reduced energy change of reaction, $\hat{\upsilon}=\upsilon/k_\text{B}T_\text{c}$ is the reduced volume change of reaction, and $\hat{s}=s/k_\text{B}$ is the entropy change of reaction.\cite{anisimov_thermodynamics_2018,Caupin_Minimal_2021} 

The reduced chemical potential for such a system undergoing spinodal decomposition towards both chemical-reaction and phase equilibrium, $\hat{\mu}=\mu/k_\text{B}T_\text{c}$, is the reduced deviation of the chemical-potential difference in solution ($\mu = \mu_A - \mu_B$) from its equilibrium value, $\mu = 0$. The reduced time-dependent chemical potential is found from the functional derivative of Eq.~(\ref{Eqn_LGfunc})
\begin{equation}\label{Eqn_MuOpen}
    \hat{\mu} = \hat{\mu}_\text{th}-\kappa\nabla^2\phi
\end{equation}
where the order parameter depends on space and time, $\phi=\phi(\vec{r},t)$. In this form, the chemical potential is comprised of a thermodynamic potential (given through the notation suggested in ref.\cite{Lamorgese_Liquid_2011}) as $\hat{\mu}_\text{th} = \partial\hat{G}/\partial\phi$, and a spatial-dependent part, $\kappa\nabla^2\phi$. The thermodynamic component is
\begin{equation}\label{Eqn_pdvGibbs}
    \hat{\mu}_\text{th} =\frac{\hat{T}}{2}\ln\left(\frac{1+\phi}{1-\phi}\right) -\frac{\epsilon}{2}\phi -\frac{1}{2}\left(\hat{e} + \hat{\upsilon}P-s\hat{T}\right)
\end{equation}
Expanding the first term to first order in $\phi$ via a Taylor series around $\phi =0$ (the value of the order parameter at the initial time $t=0$), gives $\partial \hat{G}/\partial \phi\approx \hat{\chi}^{-1}_{q=0}\phi$, where the inverse thermodynamic susceptibility in the limit of zero wavenumber is $\hat{\chi}^{-1}_{q=0}=\partial^2\hat{G}/\partial\phi^2$. In the mean-field regular-solution model, the inverse susceptibility scales as $\hat{\chi}^{-1}_{q=0}\sim \Delta\hat{T}$, where $\Delta\hat{T} = T/T_\text{c}-1$ is the reduced distance to the critical temperature. Therefore, the chemical potential defined in Eq.~(\ref{Eqn_MuOpen}) in the first-order approximation becomes
\begin{equation}\label{Eqn-ChemPot}
    \hat{\mu}\approx \Delta\hat{T}\phi -\frac{1}{2}\left(\hat{e} + \hat{\upsilon}P-s\hat{T}\right) -\kappa\nabla^2\phi
\end{equation}
We emphasize that the reduced chemical potential, given by Eq.~(\ref{Eqn-ChemPot}), when the term related to interconversion is absent, is the same as obtained in the classical Cahn-Hilliard theory.\cite{cahn_phase_1965}

\subsection{Kinetics of Phase Domain Growth Affected by Interconversion \label{Subsec_KineticEqns}}
The continuity equation for an incompressible binary mixture is typically given in the form \cite{Groot_NonEquilibrium_1984,carati_chemical_1997,puri_kinetics_2004,Lamorgese_Mixing_2006,Lamorgese_Liquid_2011,Mauri_NoneqThermo_2013,Lamorgese_Spinodal_2016,li_non-equilibrium_2020}
\begin{equation}\label{Eqn_Continuity}
    \pdv{\phi}{t} + \div{\vec{J}_\text{C}} = J_\text{NC}
\end{equation}
where $\vec{J}_\text{C}$ is the mutual diffusion flux associated with the conserved component of the order parameter, and it is related to the gradient of the chemical potential by $\vec{J}_\text{C}=-M\grad{\hat{\mu}}$, in which $M$ is the molecular mobility. Similarly, the flux associated with the nonconserved component of the order parameter is $J_\text{NC}$, which (for reaction-diffusion system) is expressed through the reaction rate.\cite{Lamorgese_Liquid_2011,Mauri_NoneqThermo_2013,Lamorgese_Spinodal_2016}

In Cahn-Hilliard's theory of spinodal decomposition, when there is only mutual diffusion (no interconversion of any form), the number of molecules of a particular species is conserved such that $J_\text{NC}=0$ and Eq.~(\ref{Eqn_Continuity}) reduces to the classical Cahn-Hilliard equation \cite{cahn_phase_1965}
\begin{equation}\label{Eqn_CahnHill_Diff}
    \pdv{\phi}{t} = M\laplacian{\hat{\mu}}
\end{equation}
This equations describes the dynamics of a system that will evolve toward phase separation when quenched into the unstable region (under the spinodal). This equation can be analytically solved with use of Fourier analysis to determine the phase domain growth rate\cite{cahn_phase_1965}, which, in the Cahn-Hilliard theory, is given by
\begin{equation}\label{Eqn_CahnHill_Amp}
    \omega(q) = -D q^2(1-\xi^2q^2)
\end{equation}
where $\xi$ is the correlation length of concentration fluctuations and $D$ is the mutual diffusion coefficient; in the mean field approximation, $\xi^2 = \kappa /|\hat{\chi}^{-1}_{q=0}|$ and $D = M\hat{\chi}^{-1}_{q=0}$. In the unstable (under the spinodal) region, the susceptibility is negative, the correlation length by definition is positive, and, consequently, the mutual diffusion coefficient is negative. Hence, mutual diffusion is the driving force for phase separation through spinodal decomposition \cite{cahn_phase_1965,Mauri_SpinodalDecomp_1996,Vladimirova_2DModel_1999}.

The flux, associated with the nonconserved component of the order parameter, in the continuity equation, Eq.~(\ref{Eqn_Continuity}), originates due to the two types of interconversion reactions present in the system, and consequently, $J_\text{NC}$, may be expressed through two terms,\cite{hohenberg_theory_1977}
\begin{equation}\label{Eq_Cont_Rate}
    J_\text{NC} = -L\hat{\mu}+\pi
\end{equation}
where the first term, $-L\hat{\mu}$, with an interconversion kinetic coefficient $L$, represents the rate of natural interconversion and with the chemical potential, given by Eq.~(\ref{Eqn-ChemPot}). The second term, $\pi$, denotes a nonequilibrium source of forceful interconversion that requires the alternative species to remain in equal amounts, and is analogous to the external magnetic or electric field in ferromagnetic or ferroelectric systems\cite{hohenberg_theory_1977}.

We first consider the simplest chemical reaction in which the interconversion equilibrium constant $\mathcal{K}= k_1 / k_2 = 1$, which does not depend on temperature or pressure, such that $\ln\mathcal{K} = 0$.  Therefore, the Gibbs energy change of reaction, given by Eq.~(\ref{Eq_GBA}), is $\hat{G}_{BA}=0$\cite{LL_Stat_Phys}, (we will consider the more general case, when $\mathcal{K}=\mathcal{K}(T,P)$, in Sec.~\ref{Sec_Polyamorphism}). In this approximation, the interconversion between $A$ and $B$ mirrors the flipping of spins in the Ising model, in which there is no heat or volume change of the reaction. Thus, the $-L\hat{\mu}$ term corresponding to the natural interconversion represents the nonconserved evolution of the order parameter, in which the conservation of molecules of a certain species is no longer a conserved property.\cite{cahn_microscopic_1977}



As for the source term, $\pi$, depending on the system, it can be written in a variety of forms, but typically, it characterizes the additional interactions present in a system. It may be classified as a ``source'' ($\pi>0$) or a ``sink'' ($\pi < 0$) of diffusion or interconversion, and it may exist as an equilibrium source or be introduced externally, producing a nonequilibrium system. For example, the internal source can be found in systems of diblock copolymers, where the finite size of the block competes with phase segregation, leading to the formation of equilibrium microphase domains\cite{Ohta_Block_1986,Glotzer_consistent_1994}. Another example of this type of source is in ferromagnetic or ferroelectric systems and dipolar fluids, in which the long-ranged Coulombic interactions may produce equilibrium microphase separation\cite{li_non-equilibrium_2020}. A diffusion-promoting  source could also be introduced externally through radiation to enhance or inhibit phase separation\cite{Enrique_Compositional_2001}.

Alternatively, an external interconversion source could be achieved via the interactions of energy-carrying particles, such as photons, that may break intramolecular bonds\cite{Miyata_PolyemrMix_2017} or it could be seen in biological cells through a flux of energy produced by ATP.\cite{Boeynaems_Protein_2018,Hyman_LLBiology_2014} Likewise, the source could promote interconversion through the flux of matter, such as in a system constrained to an adsorbing-desorbing layer\cite{Verdasca_Chemically_1995}. Lastly, an internal interconversion source originating from a disbalance of intermolecular forces could produce a nonequilibrium system with steady-state microphase domains. This effect was observed by Uralcan \textit{et al.}\cite{Uralcan_Interconversion_2020} in a dissipative chiral-mixture model. This case is considered in Sec.~\ref{Subsec_DissChiral}.


In this work, we consider a source of forceful interconversion that is independent of the chemical potential in the equilibrium state; one that will drive the system towards a ``completely interconverted state'' - a spatially homogeneous state with equal amounts of interconverting species - by allowing the alternative species to interconvert against the equilibrium conditions via Eq.~(\ref{Eqn-INT}). Thus, forceful interconversion may be viewed as a local chemical reaction for which the chemical potential difference depends only on the bulk concentration \cite{Lamorgese_Spinodal_2016}. For this reason, we will refer to such a source as a ``forceful interconversion source''.

If we consider the effect of a forceful interconversion on a system with conserved order-parameter dynamics ($L = 0$), this corresponds to the model originally introduced by Glotzer \textit{et al}. \cite{glotzer_monte_1994,glotzer_reaction-controlled_1995} where in the first order approximation (valid only near thermodynamic equilibrium\cite{Lamorgese_Spinodal_2016}) the forceful interconversion term in Eq.~(\ref{Eq_Cont_Rate}) has the form of $\pi = k_2c_B - k_1c_A$. Rearranging this reaction rate using the fact that the total number of particles in the system is conserved, $c_A + c_B = 1$ and $K = k_1 = k_2$, then the source of forceful interconversion may be written in terms of the order parameter ($\phi$) as $\pi = - K\phi(\vec{r},t)$. In this case, Eq. (\ref{Eqn_Continuity}) is given in the form
\begin{equation}
    \pdv{\phi}{t} = - K\phi + M\laplacian{\mu}
\end{equation} 
The corresponding growth rate for this system is $\omega(q) = -K - D q^2(1 - \xi^2q^2)$. Glotzer \textit{et al.} found that the addition of such a specific type of forceful interconversion source causes the system to phase separate into microphase domains with a characteristic length scale less than the size of the simulation box - the phenomenon of nonequilibrium microphase separation. They show that the growth rate is restricted at the wavenumber, $q_{-} = \sqrt{-K/D}$, corresponding to the first root of $\omega(q)=0$. This model describes the formation of microphase domains in a nonequilibrium steady-state in the presence of a forceful interconversion. \cite{lefever_comment_1995,carati_chemical_1997,Lamorgese_Spinodal_2016} We note that this phenomenon could be conceptually compared to the common ``household'' example of stirring a mixture of water and oil. Obviously, there is no chemical interconversion between water and oil molecules. However, the water and oil will separate into phase domains, whose size depends on the stirring rate, $K$.

It is important to note that, in the first-order approximation, the form of the source of forceful interconversion is similar to the thermodynamic part of the chemical potential, $\hat{\mu}_\text{th}$. However, as shown by Eq.~(\ref{Eqn_pdvGibbs}), $\hat{\mu}_\text{th}$ in the mean-field approximation is linearly dependent on the distance to the critical temperature, while the source only depends on the order parameter, $\phi$ and the reaction rate $K$. 

Combining Eq.~(\ref{Eq_Cont_Rate}) for $J_\text{NC}$ and the source of forceful interconversion considered by Glotzer \textit{et al.}, $\pi = -K\phi$, the continuity equation, Eq.~(\ref{Eqn_Continuity}), may be expressed in the generalized form
\begin{equation}\label{Eqn-GeneralDifferential}
	\pdv{\phi}{t} = M\laplacian{\hat{\mu}} - L\hat{\mu} - K\phi 
\end{equation} 
In which the three terms with kinetic coefficients, $M$, $L$, and $K$ represent the mutual diffusion, natural interconversion, and forceful interconversion dynamics.

It can be shown that the reduced entropy production, $\hat{\sigma}=\sigma/k_\text{B}$, obtained using the derivation from Groot and Mazur\cite{Groot_NonEquilibrium_1984} as well as by Mauri\cite{Mauri_NoneqThermo_2013}, has the form $\hat{\sigma} = -\vec{J}_\text{C}\cdot\grad{\hat{\mu}} +J_\text{NC}\hat{\mu}_\text{th}$. Using the mutual diffusion flux, $\vec{J}_\text{C}$, from the discussion of Eq.~(\ref{Eqn_Continuity}) and the interconversion flux, $J_\text{NC}$, from Eq.~(\ref{Eq_Cont_Rate}), the entropy production may be written in the simplified form
\begin{equation}
    \hat{\sigma} = M|\grad{\hat{\mu}}|^2 + L\hat{\mu}\hat{\mu}_\text{th} + K\phi\hat{\mu}_\text{th}
\end{equation}
where it can be seen that $\hat{\sigma}>0$ always, and specifically, in the equilibrium limit when, $\hat{\mu}=0$, the entropy production remains $\hat{\sigma}>0$ due to the presence of the external source of forceful interconversion.

Using the chemical potential described by Eq.~(\ref{Eqn-ChemPot}), we can express Eq.~(\ref{Eqn-GeneralDifferential}) in the expanded form
\begin{equation}\label{Eqn-DiffFinal}
	\pdv{\phi}{t} = - (K + L\hat{\chi}^{-1}_{q=0})\phi + (M\hat{\chi}^{-1}_{q=0} + L\kappa)\laplacian{\phi} - M \kappa\nabla^4{\phi}
\end{equation}
This differential equation has the following solution:
\begin{equation}\label{Eqn_AverageOrderParam}
	\phi = \phi_0+\sum_{i} \phi_\infty e^{\omega(q_i)t}\cos(\vec{q}_i\cdot\vec{r})
\end{equation}
where $\phi_0$ and $\phi_\infty$ are constants determined by the initial ($t=0$) and steady-state ($t\to\infty$) conditions of the order parameter, respectively. Also, $\omega(q)$ is the generalized growth rate, defined as
\begin{equation}\label{Eqn-R(q)_old}
	\omega(q) = -(L\hat{\chi}^{-1}_{q=0} + K) - (M \hat{\chi}^{-1}_{q=0} + \kappa L)q^2 - M\kappa q^4
\end{equation}
We note that a conceptually similar equation for the growth rate in a system with an auto-catalytic reaction\cite{huberman_striations_1976} was also obtained by Lefever \textit{et al.} \cite{carati_chemical_1997} Eq.~(\ref{Eqn-R(q)_old}) can also be expressed through the susceptibility, $\hat{\chi}_{q=0}$, and the correlation length, $\xi$, as
\begin{equation}\label{Eqn-R(q)}
    \omega(q) = - K -\hat{\chi}^{-1}_{q=0}(Mq^2+L\kappa)(1-\xi^2q^2)
\end{equation}
The phase domain growth rate describes the characteristics of phase domain growth for both phase amplification and microphase separation.

\subsection{Phase Amplification \textit{vs.} Microphase Separation \label{Subsec_PhaseAmpVsMicSep}}

At this point, we will discuss the key differences between phase amplification, microphase separation, and the traditional Cahn-Hilliard phase segregation without interconversion, which we refer to as ``complete phase separation.'' First, we emphasize that the phase transition through phase amplification is fundamentally different from phase separation. Phase amplification occurs to avoid the formation of an energetically unfavorable interface between alternative stable phase domains. This phenomenon is only possible due to the nonconserved nature of the order parameter. In contrast, in a phase separating binary mixture the formation of an interface is required due to the conserved nature of the order parameter. However, we note that in macroscopic systems where the interfacial energy is much smaller than the bulk energy, a system with a nonconserved order parameter may enter a metastable state in which an interface forms between phases\cite{Mazenko_Theory_1984,Marko_Phase_1995,Shum_Phase_2021}. Thus, the size of the system plays a crucial role in phase amplification. With increasing system size, the energy of the surface decreases when compared to the bulk energy, the conformational energy of the metastable interface becomes less unfavorable, and the possibility that the system will form an interface drastically increases.

\begin{table}[b]
    \centering
    \caption{Limiting cases of the interplay between diffusion, natural interconversion, and forceful interconversion.}
    \begin{tabular}{c|c}
    Conditions & Phenomenon           \\\hline
    & \\
    \makecell{ Only Diffusion \\ ($L=0$ \& $K=0$)}  & \makecell{Complete Phase \\Separation} \\
    & \\
    
    \makecell{ Only Natural \\ Interconversion \\ ($M=0$ \& $K=0$)} & \makecell{Unrestricted Phase \\Amplification}\\
    & \\
    
    \makecell{ Only Forceful \\ Interconversion \\ ($M=0$ \& $L=0$)}  & \makecell{Homogeneous \\Steady State}
    \end{tabular}
    \label{Table_LimitingCases}
\end{table}

Second, similar to phase amplification, the size of the system and the rate of forceful interconversion are important for microphase separation to occur. As we will discuss in Sec.~\ref{Subsec_DissChiral2}, we have found that there are two key conditions required to observe microphase separation. First, if the characteristic size of the mesoscopic steady-state microphase domains are comparable to half the size of the system, then the system will produce the same two alternative phases that would be observed without interconversion. As a result, the size of the system may ``cut off'' the system’s ability to phase separate into microdomains. Second, if the rate of forceful interconversion is much faster than the natural interconversion or diffusion rate ($Dq^2$), then the external force dominates the systems’ kinetics and no phase formation is possible, and consequently, the characteristic length scale of the microphase emerges as a result of the competition between forceful interconversion and diffusion.

\begin{figure}[t]
	\centering
	\includegraphics[width=\linewidth]{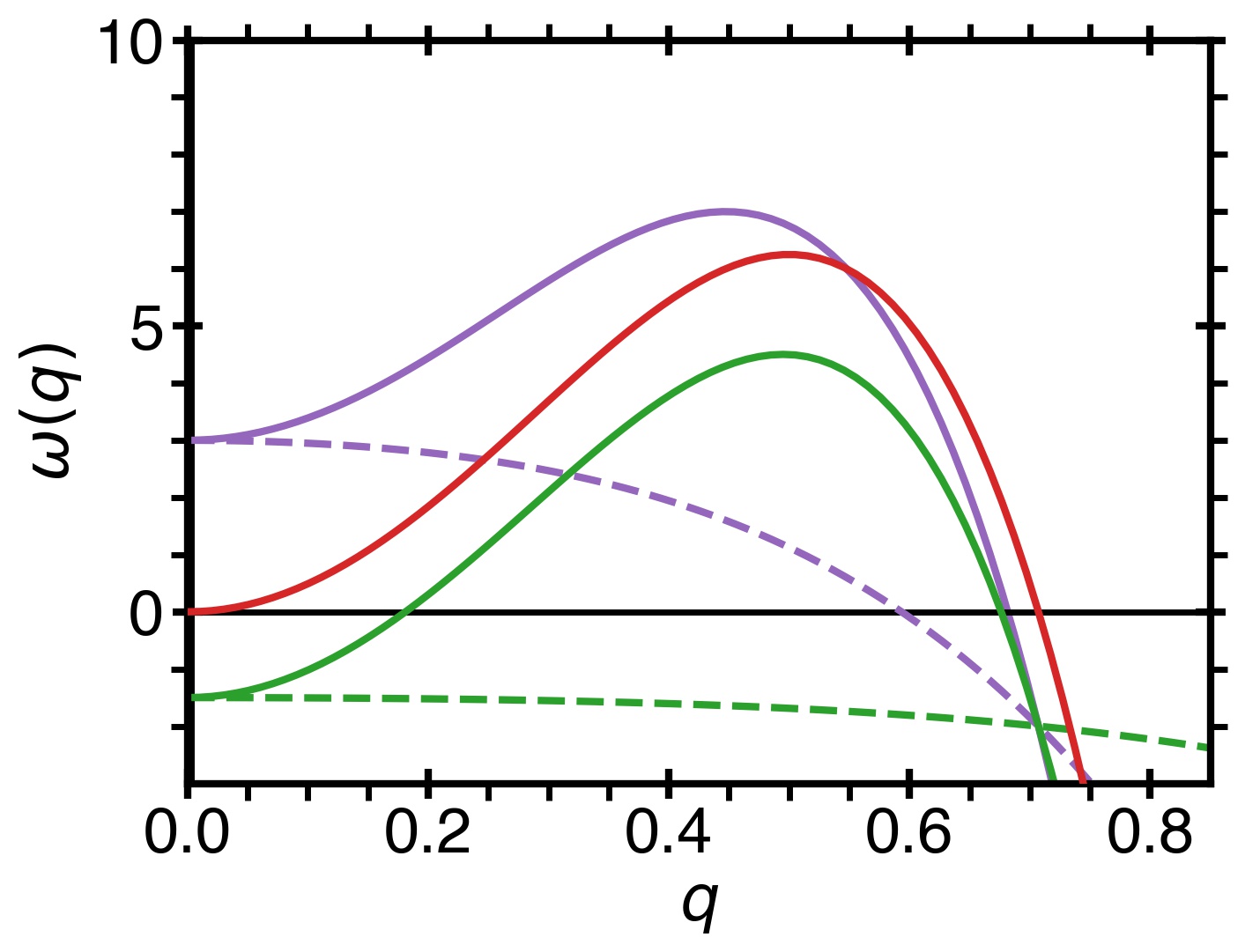}
	\caption{The characteristic growth rate, Eq. (\ref{Eqn-R(q)}), affected by the competition between diffusion, natural interconversion, and forceful interconversion at $\Delta\hat{T}=-0.5$. Complete phase separation (as predicted by Cahn-Hilliard's theory for $L=0$ and $K=0$) is illustrated by the red curve for $M = 100$. Phase amplification is illustrated by the purple curves for restricted ($M = 100$ - solid) and unrestricted ($M = 10$ - dashed) cases, in which $L = 10$ and $K = 2$.  Microphase separation for $M = 100$, $L = 1$, and $K = 2$ is illustrated by the solid green curve. When the growth rate is always negative, as illustrated by the green dashed curve (for $M = 1$, $L = 1$, and $K = 2$), there is no phase domain growth corresponding to a homogeneous steady state.}
    \label{Fig-R(q)Vsq}
\end{figure}

Whether phase amplification or microphase separation will occur depends on the interplay of the three dynamics in the system: diffusion, natural interconversion, and forceful interconversion, as given through the characteristic growth rate,  Eq. (\ref{Eqn-R(q)}). In the limiting cases when one of the rates dominates the system, complete phase separation, unrestricted phase amplification, or a homogeneous steady state will be observed. The results of these observations are summarized in Table \ref{Table_LimitingCases}. 

In a system with mixed dynamics where diffusion, natural interconversion, and forceful interconversion are present, the interplay between these three rates produces either phase amplification, microphase separation, or a homogeneous steady state with no domain growth. The resultant effect on the system may be determined through the shape and intercepts of the characteristic growth rate. For instance, the competition between natural interconversion ($L\hat{\chi}^{-1}_{q=0}$) and forceful interconversion ($K$), as follows from Eq. (\ref{Eqn-R(q)_old}), shifts the intercept of $\omega(q=0)$ up or down producing either phase amplification or microphase separation depending on their magnitude. Meanwhile, the competition between diffusion ($M\hat{\chi}^{-1}_{q=0}$) and natural interconversion ($L\kappa$) dynamics is described by the concavity and convexity of the growth rate around $q=0$, which when combined with the position of the intercept determines whether restricted or unrestricted phase amplification will be observed or if the system will remain homogeneous instead of phase separating into microphase domains. Characteristic growth rates for various relationships between $M$, $L$, and $K$ are shown in Figure~\ref{Fig-R(q)Vsq} and summarized in Table \ref{Table1}. 

To achieve phase amplification, the forceful interconversion rate must be slower than the natural interconversion rate, $K < L|\hat{\chi}^{-1}_{q=0}|$, such that the growth rate is shifted up. Alternatively, microphase separation occurs when the forceful interconversion rate and the diffusion rate is faster than the natural interconversion rate, such that $K > L|\hat{\chi}^{-1}_{q=0}|$ and $M |\hat{\chi}^{-1}_{q=0}| > \kappa L$, respectively. We note that in the case when forceful interconversion is faster than natural interconversion (while the diffusion rate is slower than the interconversion rate) the system will be in a homogeneous state as the growth rate is negative for all wavenumbers.

\begin{table}[t]
	\centering
	\caption{Conditions for phase amplification and microphase separation as illustrated in Figure~\ref{Fig-R(q)Vsq}. The left column corresponds to the solid lines and the right column corresponds to the dashed lines.}
	\begin{tabular}{c|cc}
		& $M |\hat{\chi}^{-1}_{q=0}| > L\kappa$ & $M |\hat{\chi}^{-1}_{q=0}| < L\kappa$ \\\hline
    	&                        & \\
		$L |\hat{\chi}^{-1}_{q=0}| > K$ &  \makecell{Restricted (Slow) \\Phase Amplification}       & \makecell{Unrestricted (Fast)\\ Phase Amplification}\\
		&                        & \\
		$L |\hat{\chi}^{-1}_{q=0}| < K$ & \makecell{Microphase \\Separation}    & \makecell{Homogeneous\\ Steady State}
	\end{tabular}
	\label{Table1}
\end{table}

The conditions for whether the system will achieve complete macro phase separation, undergo microphase separation, or experience phase amplification can be determined from the characteristic wavenumbers of the growth rate, Eq. (\ref{Eqn-R(q)}), found from its two roots and maximum. These wavenumbers are: $q_m^\omega$, the wavenumber corresponding to the fastest growing inhomogeneities; $q_{-}$, the first root of $\omega(q)$; and $q_{+}$, the second root of $\omega(q)$. These three characteristic wavenumbers are related through $q_{-}^2 = 2(q_m^\omega)^2 - q_{+}^2$. As will be proven in Sec. \ref{Sec-Sfactor}, the existence of a non-zero $q_{-}$ indicates steady-state microphase separation; such that, after infinite time, the size of the steady-state phase domain, $R_\infty$, will be described by $R_\infty \sim 1/q_{-}$, which in the lowest order approximation is given through $M$, $L$, and $K$ as
\begin{equation}\label{Eqn_Minusqn}
    q_{-}^2 = \frac{K-L\hat{\chi}^{-1}_{q=0}}{M\hat{\chi}^{-1}_{q=0}-L\kappa} = \frac{K-L\hat{\chi}^{-1}_{q=0}}{D_\text{eff}}
\end{equation}
where $D_\text{eff} = -\hat{\chi}^{-1}_{q=0}(M-L\xi^2)$ is the effective mutual diffusion coefficient modified by interconversion kinetics. Additionally, solving Eq.~(\ref{Eqn-R(q)}) for $q_m^\omega$, gives
\begin{equation}\label{Eqn_Maxqn}
    (q_m^\omega)^2 = \frac{1}{2\xi^2}\left(1-\frac{L}{M}\xi^2\right)
\end{equation}
The maximum of the growth rate is shifted in comparison to Cahn-Hilliard theory, $q_m^\omega = 1/(\sqrt{2}\xi)$,\cite{cahn_phase_1965} as shown in Figure~\ref{Fig-R(q)Vsq}. This shift is independent of the strength of the source of forceful interconversion.

\section{\label{Sec-Sfactor} Structure Factor: Manifestation of Microphase Separation}
We have observed in atomistic models\cite{Uralcan_Interconversion_2020,Shum_Phase_2021,Glotzer_consistent_1994,glotzer_reaction-controlled_1995} (discussed in Sec. \ref{Sec_Apps}) and through computational calculations of the time-dependent structure factor for the system\cite{Longo_Structure_2021} that the steady-state domain size may be predicted from the lower cut-off wave number, $q_-$, determined from the growth rate, Eq.~(\ref{Eqn-R(q)}). In this section, we describe the generalized Cahn-Hilliard-Cook structure factor which includes both natural and forceful interconversion of species\cite{Longo_Structure_2021}, and using a simple crossover function, we incorporate the evolution from the spinodal regime to the nucleation regime.

\subsection{Generalized Cahn-Hilliard-Cook Structure Factor}
The time-dependent structure factor, $S(q,t)$, is given through the integral of the correlation function for the concentration fluctuations \cite{Bray_Theory_2002}. The equation of motion for the structure factor is found by introducing order parameter fluctuations, $\delta\phi(\vec{r},t)$, into the general equation for the time evolution of the order parameter, Eq.~(\ref{Eqn-GeneralDifferential}), and spatially integrating $<\delta \phi(\vec{r},t)\delta \phi(\vec{r}_0,t)>$\cite{cook_brownian_1970,langer_new_1975}. Doing so, the first-order solution for mixed diffusion-interconversion dynamics is\cite{Longo_Structure_2021}
\begin{equation}
	\pdv{S(q,t)}{t} = 2\omega(q,t)S(q,t) + 2\left(Mq^2 + L\kappa\right)
\end{equation}
where $\omega(q,t)$ is given by Eq.~(\ref{Eqn-R(q)}) \cite{Glotzer_consistent_1994,Coniglio_Multiscaling_1989,Coniglio_Novel_1990}. We note that in the absence of natural interconversion and forceful interconversion, this equation reduces to the result presented by Cook \cite{cook_brownian_1970}. Solving this differential equation for the structure factor, assuming a linear approximation \cite{Langer_Theory_1973}, $\partial\omega/\partial t \ll \omega(q,t)$, \cite{Billotet_Dynamic_1980,Binder_Theory_1978,binder_collective_1983,Stobl_Structure_1985} and applying the boundary conditions: at $t=0$, the system is quenched from a sufficiently high temperature such that $S(q,t=0)=0$, and when $t\to\infty$, the structure factor becomes time independent, $S(q,t\to\infty)=S_\infty(q)$, then the $S(q,t)$ may be written in the form
\begin{equation}\label{Eqn-SCH}
	S(q,t) = S_{\infty}(q)\left(1 - e^{2\omega(q,t)t}\right) 
\end{equation}
which is valid from the initial stages of spinodal decomposition to the coarsening regime \cite{binder_collective_1983,Cahn_Later_1966}. The steady-state (infinite time) structure factor is given by
\begin{equation}\label{Eq-Schi}
	S(q,t\to\infty) = S_\infty(q) = \frac{Mq^2 + L\kappa}{-\omega(q,t\to\infty)}
\end{equation}
It can be seen that in the absence of forceful interconversion ($K=0$, equilibrium conditions) when either $L=0$ or $M=0$, then this equation reduces to the Ornstein-Zernike structure factor - $S_{OZ} =  \xi^2/(1+\xi^2q^2)$. To highlight the time-dependent effects on the growth rate, it is most convenient to express Eq.~(\ref{Eqn-R(q)}) through the wavenumbers corresponding to the maximum ($q_m^\omega$) and lower cut-off ($q_-$) of the growth rate, Eqs.~(\ref{Eqn_Minusqn} and \ref{Eqn_Maxqn}) respectively, as
\begin{equation}
    \omega(q,t) = M\kappa (q_m^\omega)^2(t) [(q_m^\omega)^2(t) -2q_-^2]-M\kappa [q^2-(q_m^\omega)^2(t)]^2
\end{equation}
In this form, the growth rate highlights the time dependence of $q_m^\omega=q_m^\omega(t)$, while illustrating that $q_-$ is an intrinsic property of the system which determines the steady-state cut-off of the growing domain sizes, $q_m^\omega(t\to\infty)\propto q_-$. The origin of the time dependence of $q_m^\omega$ is due to the change in concentration at constant temperature from the unstable ($\phi=0$) to the stable ($\phi > 0$) regime present in the higher order terms of Eq.~(\ref{Eqn_pdvGibbs}), $\hat{\chi}^{-1}_{q=0} \simeq \Delta\hat{T} + \phi^2(t)$. Based on observations in simulations \cite{Uralcan_Interconversion_2020,Longo_Structure_2021}, the time dependence of $q_m^\omega(t)$, through $\hat{\chi}^{-1}_{q=0}(t)$, may be approximated by the interpolation between the two physical limits of $q_m^\omega(t=0)$ and $q_m^\omega(t\to\infty)\to q_-$ as
\begin{equation}\label{Eq-tdepfdp}
    q_m^\omega(t) = q_m^\omega(t=0)e^{-t/\tau} + q_-(1-e^{-t/\tau})
\end{equation}
Where $\tau$ is a system-dependent parameter, which controls the crossover between spinodal decomposition and nucleation regimes.

\begin{figure}[t]
	\centering
	\includegraphics[width=\linewidth]{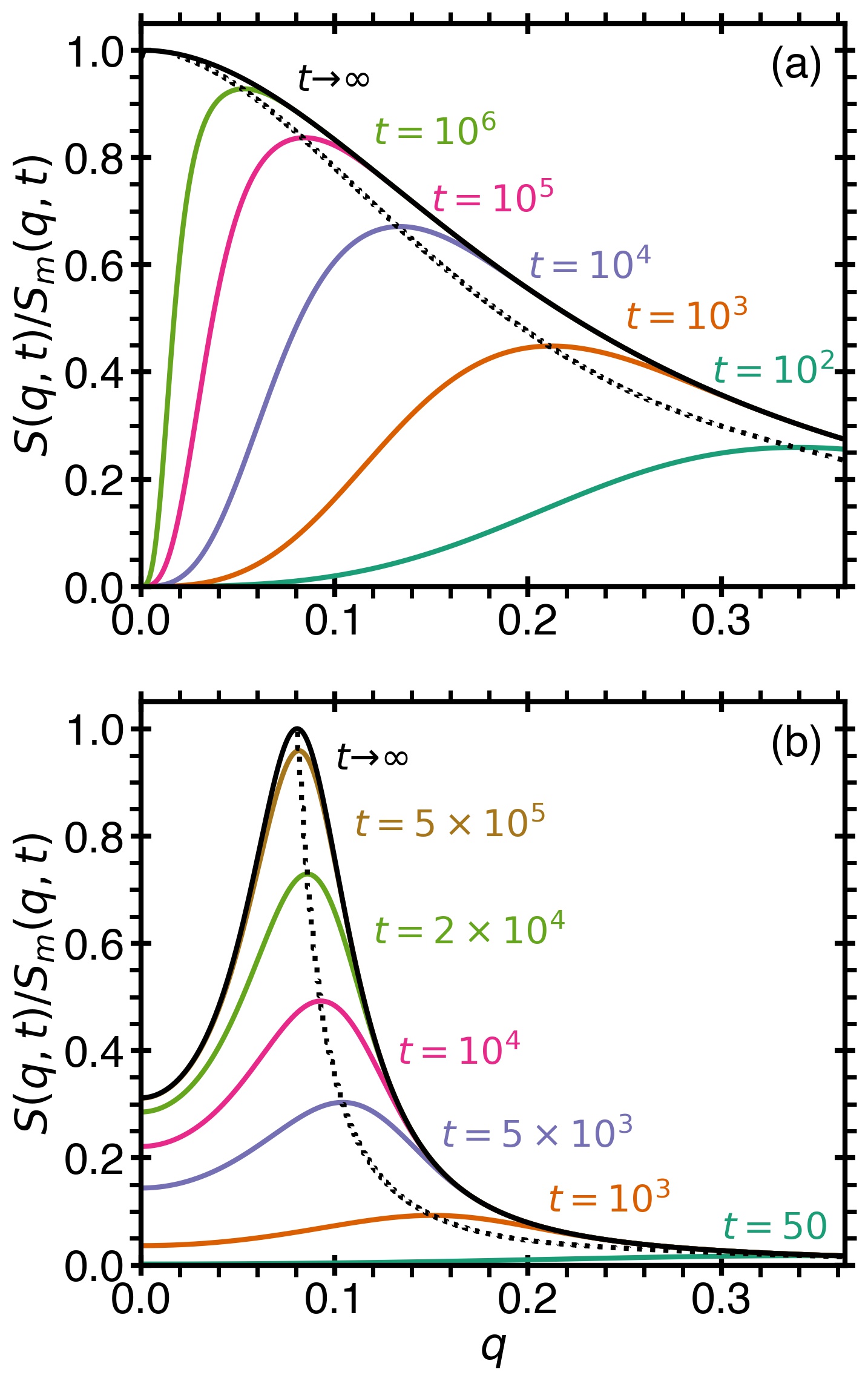}
	\caption{Temporal evolution of the structure factor: a) for a system undergoing diffusion dynamics ($M=1$) toward an equilibrium state in the absence of natural interconversion ($L=0$) and forceful interconversion ($K=0$); b) for a system undergoing a hybrid of diffusion ($M=1$) and natural interconversion ($L=0.01$) dynamics in the presence of forceful interconversion ($K=\num{1.5e-3}$) toward a steady state. The structure factor, given by Eq. (\ref{Eqn-SQTFinal}), exhibits a crossover from spinodal decomposition to the nucleation regime. The dashed-black curves indicate the development of the maximum of the structure factor. The characteristic crossover time is defined in Eq. (\ref{Eq-tdepfdp}) and adopted as $\tau = 100$. In (a) the evolution of the maximum of the structure factor moves to $q=0$ for infinite-size system and saturates at $S_\text{m}(q=0,t\to\infty) = 1/(2\xi^2) = 5$ for $\Delta \hat{T} = -0.1$. In contrast, in (b) the evolution of the maximum is interrupted at a characteristic cut-off wavenumber predicted by the characteristic phase domain growth rate, $q_m^\text{S}(t\to\infty) \propto q_-$, and it saturates at $S_\text{m}(q_-,t\to\infty)=417$.}
	\label{Fig-SqtCHSD}
\end{figure}

To describe the crossover of the time-dependent structure factor, Eq. (\ref{Eqn-SCH}), between the two regimes, the appropriate scaling behavior must still be obeyed. As given by Lifshitz-Slyozov's theory for Oswald ripening\cite{LifshitzSlurzov_Nuc,Wager_Nuc}, the nucleation regime is characterized by an Ornstein-Zernike structure factor that scales with time as $S(q,t) \propto t$ and a phase domain size that grows as $q \propto t^{1/3}$. \cite{Binder_Theory_1978} We note that in the classical Cahn-Hilliard theory for the early stage of spinodal decomposition, the phase domain size grows as $q \propto t^{1/4}$.\cite{Cahn_Later_1966} To appropriately introduce the scaling crossover from $q\propto t^{1/4}$ to $q\propto t^{1/3}$, we adopt the scaling relation for the ``crossover time'' ($t_{\cross}$), suggested by Binder \textit{et al.}, that in the nucleation regime the exponential growth scales with time and wavenumber as $\omega(q,t_{\cross}) = tq$. \cite{Binder_Theory_1978} Accordingly, we use the following Pad\'e approximant in the exponential factor in Eq. (\ref{Eqn-SCH}), such that the time, $t$, is replaced by the crossover time, $t_{\cross}$, given by
\begin{equation}\label{Eqn_Tcross}
	t_{\cross} = \frac{t(t+\tau)}{1 + tq}
\end{equation}
where $\tau$ is the same crossover parameter used for $q_m^\omega(t)$, Eq. (\ref{Eq-tdepfdp}). Lastly, to satisfy the infinite time (steady-state) limit that the structure becomes the Ornstein-Zernike structure factor, a simple crossover structure factor given by
\begin{equation}\label{Eqn-SQTFinal}
\hat{S}(q,t) = a_0S_{OZ}(q)\left(\frac{S(q,t_{\cross})}{S_{OZ}(q) +  S(q,t_{\cross})}\right)
\end{equation}
where $S(q,t_{\cross}) = S_\infty(q)\left(1 - e^{\omega(q)t_{\cross} }\right)$ and $a_0$ is a system-dependent constant. In this form, Eq.~(\ref{Eqn-SQTFinal}) represents the full crossover time-dependent structure factor from the spinodal regime to the nucleation regime.

\subsection{Cut-Off Length Scale in Microphase Separation}
Using Eq. (\ref{Eqn-SQTFinal}), we compare a system of diffusion dynamics toward equilibrium with a system of diffusion dynamics in the presence of a source of forceful interconversion toward steady state. The behavior of these systems through all three regimes, namely spinodal decomposition, coarsening, and nucleation, is shown in Figs. (\ref{Fig-SqtCHSD}a,b). It is observed that the introduction of a forceful interconversion causes the growth of the structure factor to be interrupted at the lower cut-off wavenumber, $q_-$. In the first order approximation, the wavenumber corresponding to the maximum of structure factor, $q_m^s$, in the steady-state limit of $q_m^s(t\to\infty)$ determined from $\partial S(q,t\to\infty)/\partial q = 0$ is given by
\begin{equation}\label{Eqn_qmS_value}
    q_m^s(t\to\infty) =  2^{1/4}q_- \propto \sqrt{\frac{K}{-D_\text{eff}}}
\end{equation}
We note that the general scaling law $q_m^s\propto K^{1/2}$ was observed in simulations of a chiral model\cite{Uralcan_Interconversion_2020} and a hybrid Ising/lattice-gas model\cite{Longo_Structure_2021} to be discussed in detail in Sec. \ref{Sec_Apps}. This behavior differs from studies of the microphase domain formation in block copolymers where it was found that $q_m^s\propto K^{1/4}$. \cite{Glotzer_consistent_1994,Christensen_Segregation_1996} The difference in this behavior between our binary mixture and the block copolymers might be attributed to the difference in the nature of the order parameter for the two systems - our binary mixture being described by a single component order parameter and the block copolymers being described by an $n$-component order parameter.

\begin{figure}[t]
	\centering
	\includegraphics[width=\linewidth]{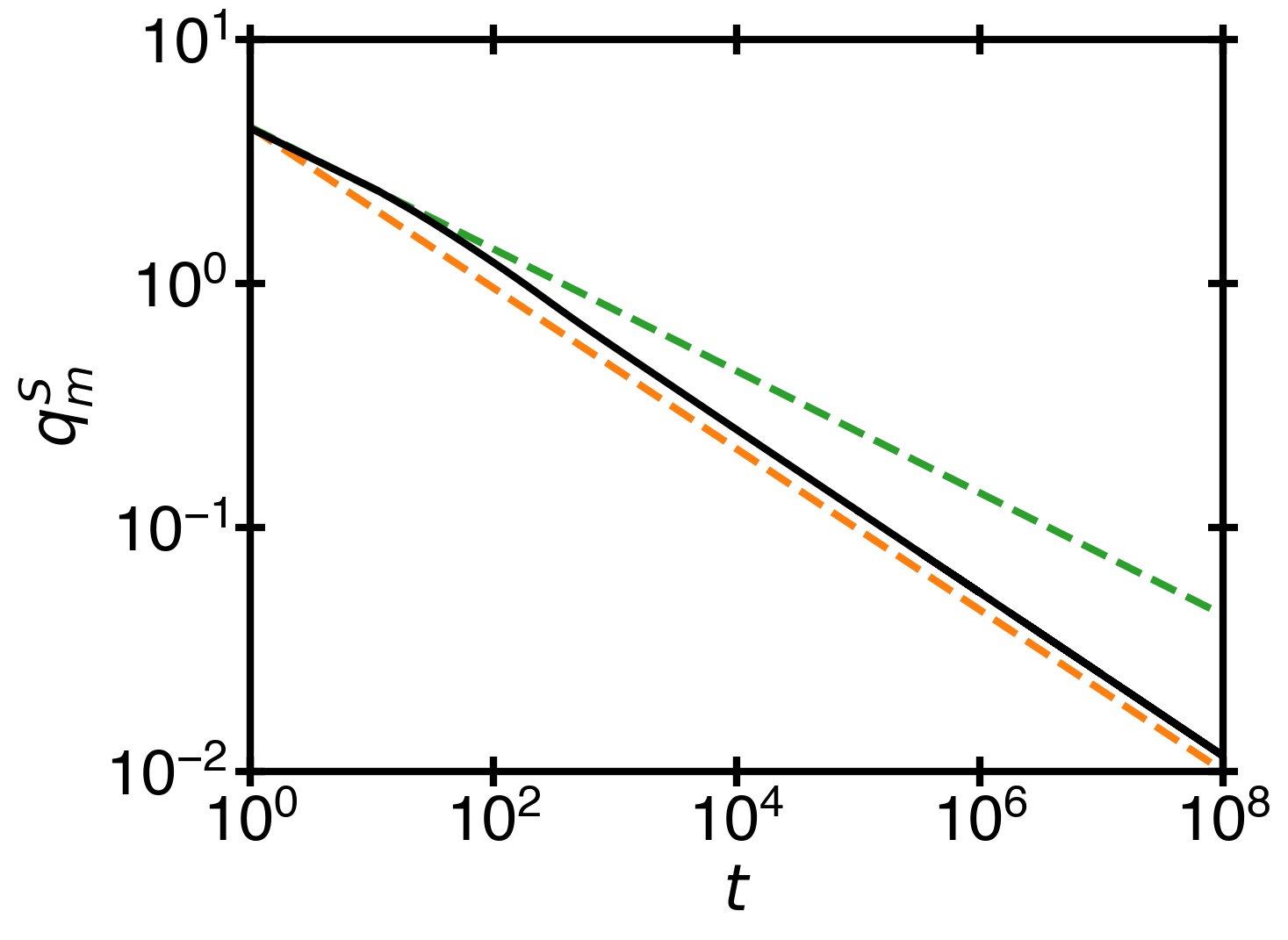}
	\caption{The time evolution of the wavenumber corresponding to the maximum of the structure factor, given by Eq. (\ref{Eqn-SQTFinal}), during the crossover from the early stage of spinodal decomposition, $q\propto t^{1/4}$ (green - dashed), to the nucleation regime, $q \propto t^{1/3}$ (orange - dashed) for a system undergoing diffusion dynamics in the absence of natural interconversion ($L=0$) and forceful interconversion ($K=0$) under conditions: $M=1$, $\Delta\hat{T}=-0.1$, $\tau = 100$.}
	\label{Fig-NucleationCrossover}
\end{figure}

The crossover behavior for the transition between spinodal decomposition and nucleation in a system with diffusion dynamics toward equilibrium is shown in Figure~ \ref{Fig-NucleationCrossover}. The wavenumber corresponding to the maximum of the structure factor, given by Eq. (\ref{Eqn-SQTFinal}), evolves according to $q \propto t^{1/4}$ (short times, spinodal decompostion regime) and $q \propto t^{1/3}$ (long times, nucleation regime). The crossover from one regime to the order is illustrated by the crossing of the orange and green dashed lines in Figure~\ref{Fig-NucleationCrossover}.

\section{Applications to Atomistic Models \label{Sec_Apps}}
In this section, we apply the generalized model of spinodal decomposition affected by interconversion to the results of simulations of two atomistic models, which were reported by Shumovskyi \textit{et al.}\cite{Shum_Phase_2021} and Uralcan \textit{et al.}\cite{Uralcan_Interconversion_2020} The first model is a hybrid Ising / binary-lattice model that exhibits unrestricted (fast) phase amplification, restricted (slow) phase amplification, or complete phase separation, depending on the conditions presented in Table (\ref{Table1})\cite{Shum_Phase_2021}. The second model is a chiral-mixture considered in two formulations: a conservative-force formulation in which it exhibits phase amplification, and a dissipative-force formulation in which it demonstrates steady-state microphase separation \cite{Uralcan_Interconversion_2020}.

\subsection{Hybrid Binary-Lattice/Ising Model: Phase Amplification \label{Subsec_HybridModel}}

In 1952, Lee and Yang showed that the Ising model for an anisotropic ferromagnet and the lattice-gas model for a ﬂuid are mathematically equivalent. \cite{lee_statistical_1952} The lattice-gas model can also be formulated as an incompressible binary lattice liquid, which could be used to describe liquid-liquid phase transitions in binary mixtures.\cite{Bertrand_Peculiar_2011} However, while the lattice-gas (or binary lattice-liquid) and Ising models are equivalent in thermodynamics, they are fundamentally different in dynamics. The order parameter in the lattice-gas model (the density or concentration) is conserved, while the order parameter in the Ising model (the magnetization) is not. \cite{hohenberg_theory_1977} As a result, the density (or concentration) relaxes to equilibrium by spatial-dependent mutual diffusion, while the relaxation of the magnetization in the lowest approximation is not spatial-dependent. An important consequence of this contrast in dynamics is the difference in the equilibrium states. In the lattice gas, below the critical temperature, two equilibrium fluid phases must coexist to conserve the total number of particles (occupied cells), while in the Ising ferromagnet only one of the alternative magnetizations, positive or negative, will survive \cite{Shum_Phase_2021}. Since the interface between the two alternative magnetic phases is energetically costly, eventually, one phase will win over the other - a process known as phase amplification \cite{Shum_Phase_2021}.

The phenomenon of phase amplification was previously observed qualitatively in simulations of chiral models exhibiting interconversion of enantiomers and referred to as ``phase bullying''. \cite{Ricci_Computational_2013,Latinwo_MolecModel_2016} Quantitatively, Shumovskyi \textit{et al.}\cite{Shum_Phase_2021} studied this phenomenon through Monte Carlo simulations of a hybrid model exhibiting diffusion (Kawasaki dynamics) and interconversion (Glauber dynamics). In this section, we compare the generalized model of spinodal decomposition affected by natural interconversion with the results of Shumovskyi \textit{et al.}\cite{Shum_Phase_2021} 

We consider an Ising-like lattice system, which represents two species A and B equivalent to a mixture of two spin types, directed up and down. These species or spins may interconvert (flip up and down), but also, since their mixing is energetically unfavorable, they will mutually diffuse toward phase separation below the critical temperature. The growth-rate equation for such a hybrid system is characterized by a mixture of conserved and non-conserved order-parameter dynamics. It can be characterized from Eq.~(\ref{Eqn-GeneralDifferential}), in the absence of forceful interconversion ($\pi=0$), as
\begin{equation}\label{Eqn-R2(q)}
	\omega(q) = -\hat{\chi}^{-1}_{q=0}(L\kappa + M q^2)(1-\xi^2q^2)
\end{equation}
When $M=0$, the nonconserved order parameter grows according to Ising spin-interconversion dynamics \cite{cahn_microscopic_1977}, while when $L=0$, the conserved order parameter grows according to lattice-gas molecular-diffusive dynamics \cite{cahn_phase_1965}. From Eq. (\ref{Eqn-R2(q)}), the probability that the system will exhibit Ising-model spin interconversion is defined as $p_r = L\kappa/(Mq^2 + L\kappa)$. If $p_r=1$, the order parameter relaxes to equilibrium through unrestricted (fast) amplification to one of two alternative phases with either positive or negative order parameter. If $p_r = 0$, the order parameter exhibits complete phase separation, and if $0 < p_r < 1$, the rate of phase amplification is restricted by the interconversion probability, the distance to the critical temperature, and the system size.

\begin{figure}[t]
    \centering
    \includegraphics[width=\linewidth]{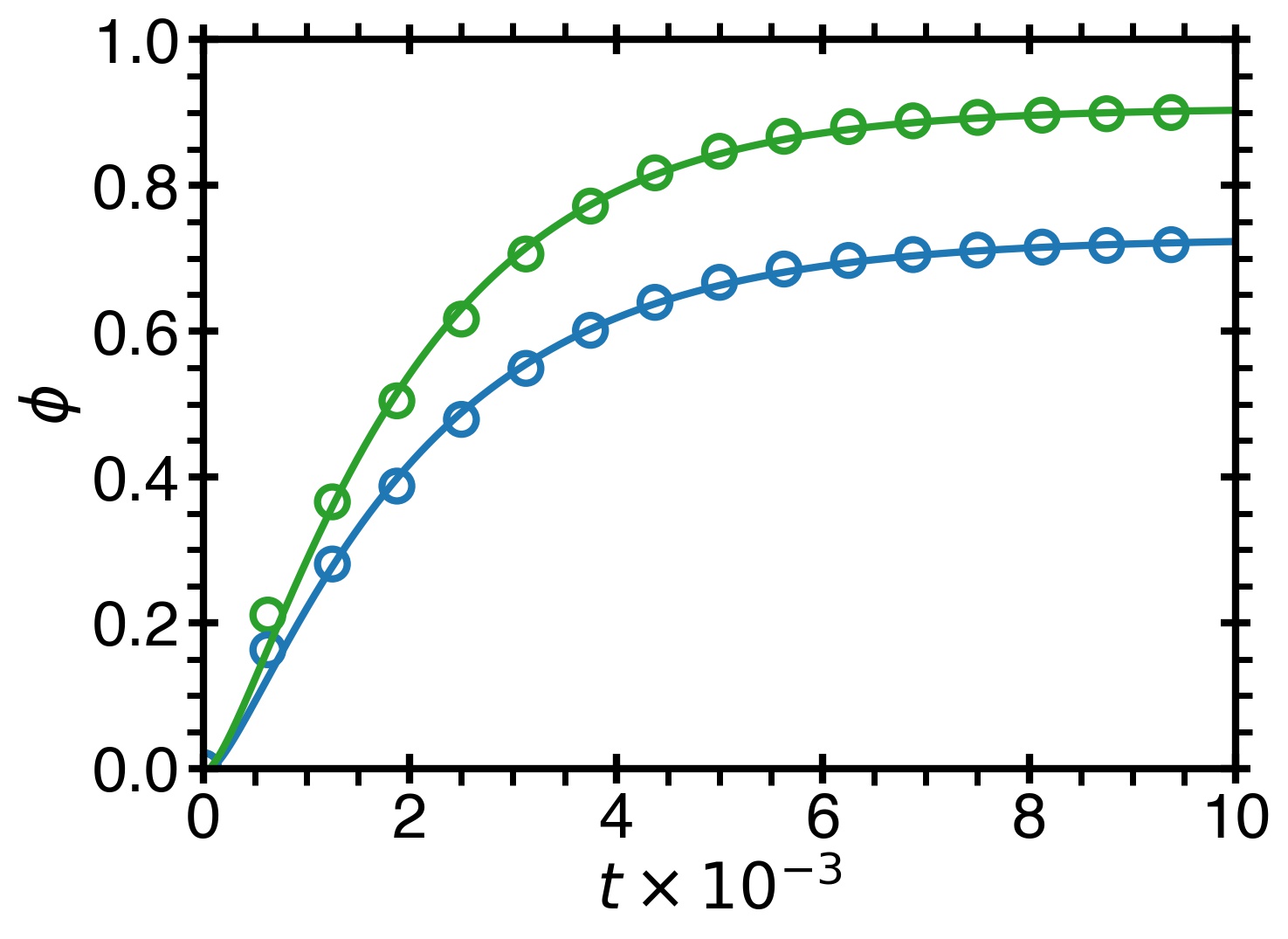}
    \caption{The time evolution of the concentration for the hybrid model with pure Ising dynamics ($p_r = 1$), given by Eq.~(\ref{Eqn_HybridConcEvo}) with the time-dependent susceptibility in the form of Eq.~(\ref{Eq_HybridSuscept}). For $\Delta\hat{T}=-0.32$ (green), $\tau = 0.2$, $\phi_1 = 0.91$, $a = 1.9$ and $b=0.3$, and for $\Delta\hat{T}=-0.11$ (blue) $\tau = 0.2$, $\phi_1 = 0.73$, $a = 4.6$, and $b=0.4$. The open circles are computational data computed from the average, absolute value of the order parameter over 100 different realizations for system size, $\ell = 100$.\cite{Shum_Phase_2021}}
    \label{Fig_Hybrid_Avg_Mag}
\end{figure}

Applying the generalized model of spinodal decomposition affected by natural interconversion, we quantify the evolution of the order parameter in the hybrid model through Eq. (\ref{Eqn_AverageOrderParam}). In a system with pure Ising dynamics ($p_r =1$), the wavenumber corresponding to the fastest growth is $q_m^\omega = 0$, hence the growth rate is $\omega(q_m)=-L\hat{\chi}^{-1}_{q=0}$. During the later stages of spinodal decomposition, as the system evolves toward nucleation, the susceptibility, at constant temperature, changes due to the time evolution of the concentration. To account for the crossover to the nucleation regime, we consider the time-dependent susceptibility evolving through the two limits: $\hat{\chi}^{-1}_{q=0}(t=0)\propto-\Delta\hat{T}$ and $\hat{\chi}^{-1}_{q=0}(t\to\infty)\propto\Delta\hat{T}$. By interpolating between these two limits, we obtain the susceptibility in the form
\begin{equation}\label{Eq_HybridSuscept}
    \hat{\chi}^{-1}_{q=0}(t)=a\hat{\chi}^{-1}_{q=0}(t=0)\left(2e^{-(t/\tau)^b}-1\right)
\end{equation}
where $a$ is a constant depending on the temperature and $b$ is an effective exponent which empirically was found to change between $b = 1/3$ to $b= 1/2$ in the interval of $\Delta\hat{T} = 0.32 - 0.06$. The temporal evolution of the order parameter as a result of phase amplification, for $p_r = 1$, is described as
\begin{equation}\label{Eqn_HybridConcEvo}
    \phi = \phi_\infty\left(1-e^{\omega(q_m^\omega)t}\right)
\end{equation}
We note that to satisfy the boundary conditions that the order parameter is $\phi =0$ at $t=0$, while also reaching its equilibrium value in the limit $t\to\infty$, then $\phi_\infty = -\phi_0/\Delta\hat{T}$, in Eq.~(\ref{Eqn_AverageOrderParam}), where $\phi_\infty$ denotes the equilibrium value of the order parameter. As illustrated in Figure~\ref{Fig_Hybrid_Avg_Mag}, the results of Shumovskyi \textit{et al.}\cite{Shum_Phase_2021} are in good agreement with this description.

We emphasize that phase amplification is manifested by the growth of the average order parameter from $\phi =0$ to its equilibrium value. In the case of pure diffusion dynamics, $p_r = 0$, the time evolution of the order parameter is described by a special form of Eq.~(\ref{Eqn_AverageOrderParam}), when the constant $\phi_0 = 0$, corresponding to the Cahn-Hilliard theory: 
\begin{equation}
    \phi = \phi_\infty e^{\omega(q_m^\omega)t}\cos(q_m^\omega r)
\end{equation}
In this case, $q_m^\omega = 1/(\sqrt{2}\xi)$, such that the average order parameter is given through the factor $\langle\cos(q_m^\omega r)\rangle = 0$. In contrast, for Ising spin-interconversion dynamics, the cosine term is evaluated at $q_m^\omega = 0$ corresponding to $\langle\cos(q_m=0\cdot r)\rangle = 1$.  Therefore, for pure diffusion dynamics ($p_r=0$), the average order parameter will remain at $\phi =0$,\cite{Shum_Phase_2021} corresponding to complete phase separation into two symmetric phases of positive and negative order parameter with the same magnitude. 

In a hybrid case, when both diffusion and interconversion are present ($0<p_r < 1$), the average of the general time evolution of the order parameter will no longer be zero, and phase amplification will be observed. This prediction is in agreement with the simulations of Shumovskyi \textit{et al.}\cite{Shum_Phase_2021}, who demonstrated that even an extremely small probability of interconversion dynamics may result in phase amplification, although the number of corresponding realizations will be exponentially small.

\subsection{Conservative-Force Formulation of the Chiral Model: Phase Amplification \label{Subsec_ConsChiral}}

The chiral model is a series of works on a 4-site tetramer model based on the simplest chiral molecule existing in nature, hydrogen peroxide\cite{Lombardo_ThermoMechanism_2009,Latinwo_MolecModel_2016,Petsev_Effect_2021,Uralcan_Interconversion_2020}. Molecular Dynamics simulations of the interconversion between the two stable enantiomeric forms is controlled through rotations around the dihedral angle. The ease at which molecules may interconvert is given through a rigidity (``spring-like'') constant $k_d$, which is related to the interconversion probability in the hybrid model through $k_d\propto \sqrt{(1/p_r)-1}$. In this model, the parameters $k_d$ and $T$ are represented in dimensionless form.\cite{Uralcan_Interconversion_2020} There are two formulations of the chiral model: a conservative-force formulation, in which all of the forces are balanced\cite{Petsev_Effect_2021} and a dissipative-force formulation, in which an imbalance of intermolecular forces is imposed between substituents of opposite chirality \cite{Uralcan_Interconversion_2020}

\begin{figure}[t]
    \centering
    \includegraphics[width=\linewidth]{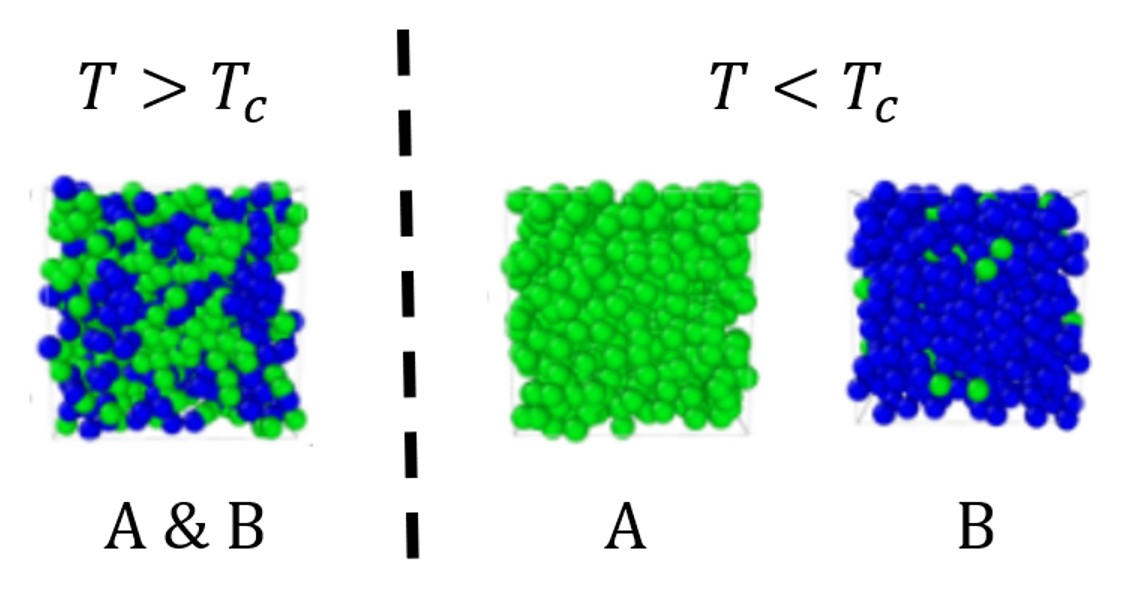}
    \caption{Phase amplification as observed in the conservative force formulation of the chiral model\cite{Uralcan_Interconversion_2020}. Above the critical temperature, the system is homogeneous, where the only apparent inhomogeneities are attributed to the correlation length of concentration fluctuations, while below the critical temperature each phase, composed of either A-type (green) or B-type (blue) enantiomers, has an equal probability of growing at the expense of the other one.}
    \label{Fig_AmpFactor}
\end{figure}

The kinetics of diffusion and interconversion for both formulations of the chiral model have been reported by Uralcan \textit{et al.}\cite{Uralcan_Interconversion_2020} In the conservative formulation, for $k_d = 10^{-3}$ (corresponding to $p_r\approx 1$ in the hybrid model), phase amplification has been observed.\cite{Uralcan_Interconversion_2020} Figure~\ref{Fig_AmpFactor} depicts snapshots of simulations of the chiral model above and below the critical temperature. Phase amplification to either A-type enantiomers, green, or B-type enantiomers, blue, will occur with equal probability.

\subsection{Dissipative-Force Formulation of the Chiral Model: Microphase Separation \label{Subsec_DissChiral}}

In this formulation of the chiral model, an imbalance of intermolecular forces produces an imbalance in chemical potential associated with the interconversion dynamics\cite{Petsev_Effect_2021}, such that the order parameter, instead of being described by Eq.~(\ref{Eqn-GeneralDifferential}) (with $\pi = 0$), evolves according to
\begin{equation}
    \pdv{\phi}{t} = M\nabla^2\mu - L\Tilde{\mu}
\end{equation}
This imbalance corresponds to the compensation of the contribution from the enthalpy of mixing in the chemical potential coupled with the interconversion kinetic coefficient, $L$. 
We note that the first term in Eq.~(\ref{Eqn-ChemPot}) could also be written through the derivatives of the reduced entropy of mixing, $\Delta \hat{S}_\text{mix}=\Delta S_\text{mix}/\rho_\text{c}k_\text{B}T_\text{c}$, and the reduced enthalpy of mixing, $\Delta \hat{H}_\text{mix}=\Delta H_\text{mix}/\rho_\text{c}k_\text{B}T_\text{c}$, as
\begin{equation}\label{Eqn_EntroEnthal}
    \Delta\hat{T}\phi = - T\pdv{(\Delta \hat{S}_\text{mix})}{\phi} + \pdv{(\Delta \hat{H}_\text{mix})}{\phi} \approx \hat{T}\phi-\phi
\end{equation}
Therefore, the racemization of enantiomers is equivalent to forceful interconversion driven only by the entropy of mixing as seen through the comparison with Eq.~(\ref{Eqn_EntroEnthal}). As a result, the ``unbalanced'' chemical potential, $\Tilde{\mu}$, is given by
\begin{equation}\label{Eqn_ChiralMu}
    \Tilde{\mu}\approx \hat{T}\phi-\kappa\nabla^2\phi
\end{equation}
Alternatively, we may attribute the effect of the energy dissipation in this model to the forceful interconversion source, $\pi$, in Eq.~(\ref{Eqn-GeneralDifferential}). In this case, $\pi$ has the form $\pi = L(\mu - \Tilde{\mu}) = -(\epsilon/2)\phi$.

\begin{figure}[t]
    \centering
    \includegraphics[width=\linewidth]{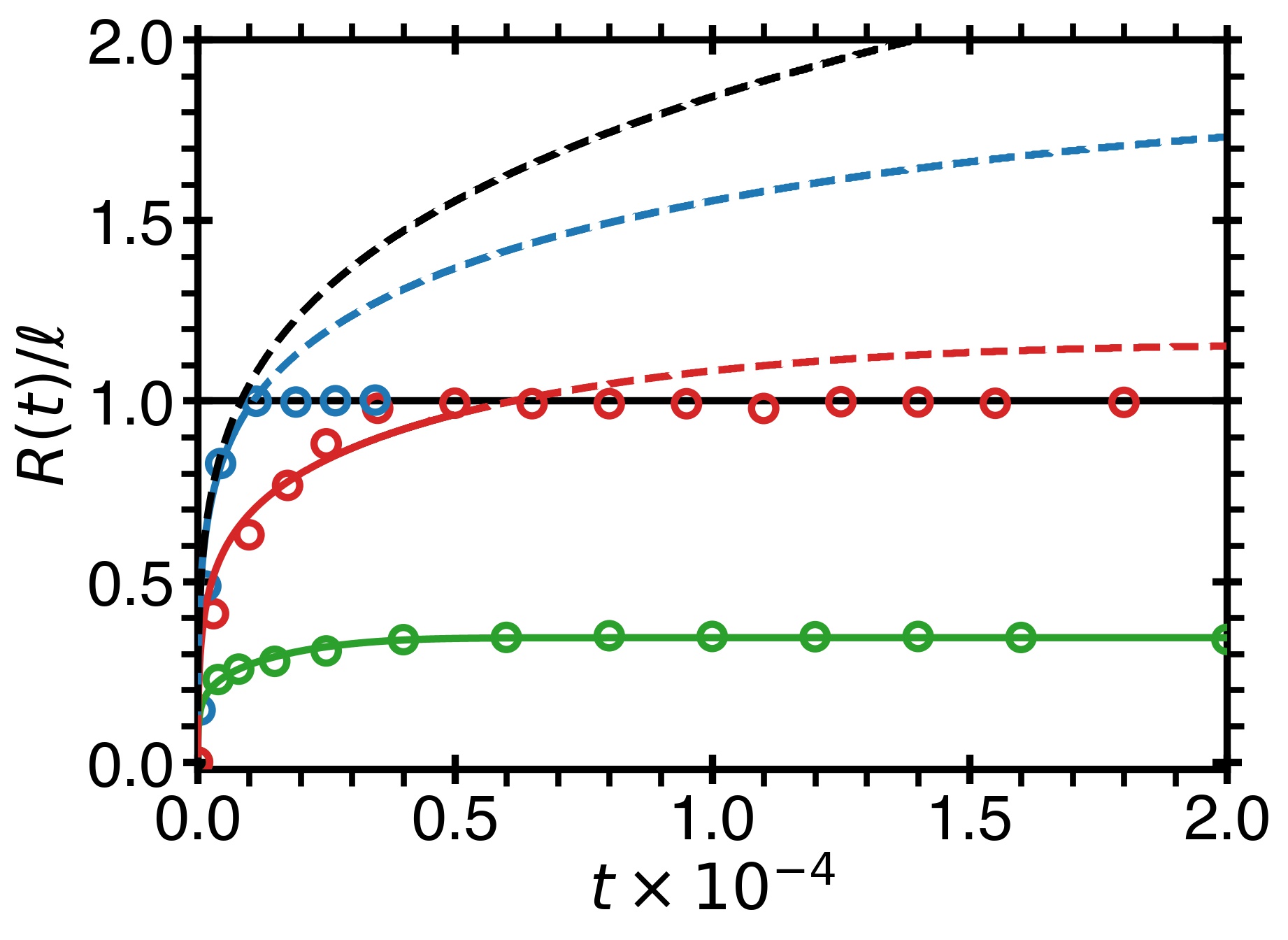}
    \caption{Phase domain growth for the dissipative-force formulation of the chiral model. The open circles represent computational data\cite{Uralcan_Interconversion_2020}, and the curves illustrates predictions of the time evolution of the domain size, $R(t) = 1/q_m^\text{S}$, from Eq.~(\ref{Eqn-SQTFinal}) for dihedral force constants: $k_d = 5$ (green), $k_d = 9.86$ (red), $k_d=19.86$ (blue), and $k_d\to\infty$ (black) for $T=1.8$, $T_\text{c}=2.3$, $\tau=2$, and $M=0.0022$. The domain size is normalized by the size of the computational box, $\ell$. The dashed curves represent the predictions of the domain growth if it is not restricted by the finite size.}
    \label{fig:DomainGrowth}
\end{figure}

The characteristic phase domain growth rate for the dissipative chiral system with a chemical potential given by Eq. (\ref{Eqn_ChiralMu}) reads
\begin{equation}\label{Eqn_ChiralAmp}
    {\omega}(q) = -L(\hat{T}+\kappa q^2) -M\Delta\hat{T}q^2(1-\xi^2q^2)
\end{equation}
Solving Eq.~(\ref{Eqn_ChiralAmp}) for $\omega(q)=0$, we obtain the first root, $q_{-}$, corresponding to the minimum wavenumber below which the growth rate becomes negative and phase domain growth will not be possible. In the first approximation, it reads as
\begin{equation}\label{Eq-qcmin_first_order}
    q_{-}^{2} = \frac{L}{D_\text{eff}}
\end{equation}
where the effective mutual diffusion coefficient is defined as $D_\text{eff} = -(\Delta\hat{T}/\hat{T})(M + L\xi^2)$ in which $\Delta\hat{T}/\hat{T} = (T-T_\text{c})/T$. The effective mutual diffusion coefficient represents the diffusion of species affected by interconversion. Remarkably, forceful interconversion of species enhances the translational molecular mobility $M$, into the effective mobility, $M_\text{eff} = M + L\xi^2$. This is a novel phenomenon that was recently discovered.\cite{Uralcan_Interconversion_2020}

As interconversion between enantiomorphs depends on the ease at which the substituents may rotate around the dihedral angle, the interconversion kinetic coefficient is assumed to be proportional to the mobility, $M$, and, as observed by Uralcan \textit{et al.}\cite{Uralcan_Interconversion_2020}, has the following dependence on the rigidity parameter, $k_d$,
\begin{equation}\label{Eqn_Chiral_INter}
    L = M \frac{T^2}{k_d^2}\left(1+a\frac{T^2}{k_d^2}\right)
\end{equation}
where $a$ is constant. For the rigidity parameters in the range investigated by Uralcan \textit{et al.}\cite{Uralcan_Interconversion_2020} ($5 < k_d < 30$), we consider the first order approximation, $L\approx MT^2/k_d^2$. We predict that the effective mobility increases with lower values of $k_d$, as $M_\text{eff} = M(1 + T^2/k_d^2\Delta\hat{T})$. This effect was indeed observed in simulations reported by Uralcan \textit{et al.} \cite{Uralcan_Interconversion_2020} Moreover, computational data on the effective diffusion rate and interconversion rate show quantitative agreement with the theoretical predictions for $D_\text{eff}$ and $L$.\cite{Uralcan_Interconversion_2020}

\begin{figure}[t]
    \centering
    \includegraphics[width=\linewidth]{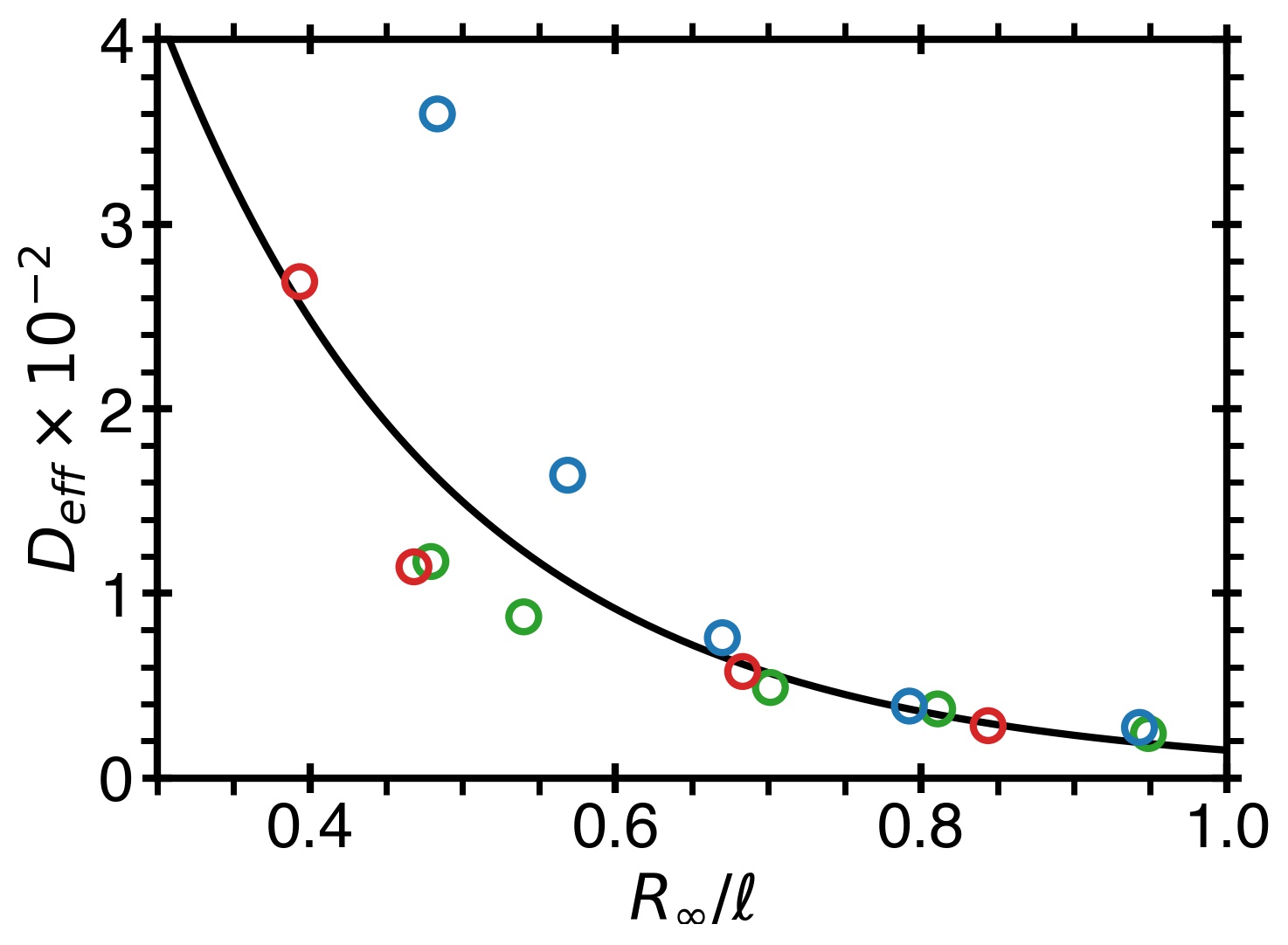}
    \caption{The mutual diffusion coefficient affected by interconversion, given by Eq.~(\ref{Eq-qcmin_first_order}), as a function of the normalized steady-state domain size $R_\infty$, Eq.~(\ref{Eqn_Chiral_SSSize}). The open circles are computational data for the dissipative-force formulation of the chiral model at three different dihedral force constants: $k_d = 5$ (green), $k_d = 9.86$ (red), and $k_d=19.86$ (blue) \cite{Uralcan_Interconversion_2020}. The molecular mobility in the limit $k_d\to\infty$ is approximated as $M = aT/\eta$, where $a = 0.94$ and $\eta$ is the viscosity approximated by the Arrhenius equation $\eta \sim e^{T_0/T}$. The characteristic temperature $T_0$ is $3.5$.}
    \label{Fig_Diffusion_Size}
\end{figure}

The phase domain growth for the dissipative-force formulation of the chiral model is illustrated in Figure~\ref{fig:DomainGrowth} for four different dihedral force constants. The domain size is calculated from the inverse wavenumber corresponding to the maximum of the time-dependent structure factor, $R(t)\propto 1/q_m^s$, as given by Eq.~(\ref{Eqn_qmS_value}). It was observed by Uralcan \textit{et al.}\cite{Uralcan_Interconversion_2020} that the steady-state domain size is inversely proportional to $k_d$, which agrees with the prediction given by Eq.~(\ref{Eqn_qmS_value}) when $k_d$ is large, $L = MT^2/k_d^2$ and $R_\infty \propto k_d$ as seen in Figure~\ref{fig:DomainGrowth}.

As a result, the characteristic size of the steady-state domains is presented in the first approximation of Eq.~(\ref{Eq-qcmin_first_order}) by
\begin{equation}\label{Eqn_Chiral_SSSize}
    R_\infty \approx \frac{k_d}{T}\sqrt{\frac{-\Delta\hat{T}}{\hat{T}}}
\end{equation}
In Figure~\ref{Fig_Diffusion_Size}, we demonstrate the agreement of the effect of forceful interconversion on the effective diffusion coefficient, given by Eq.~(\ref{Eq-qcmin_first_order}), with computational data\cite{Uralcan_Interconversion_2020}.

\subsection{Dissipative-Force Formulation of the Chiral Model: Microphase Domain Growth Restricted by Finite Size \label{Subsec_DissChiral2}}
Within the dissipative-force formulation of the chiral model, the phase domain growth is restricted by the finite size of the system $\ell$. The event when the domain size reaches the size of the computational box, was defined as ``complete'' phase separation, $\ell \sim 1/q^*$.\cite{Uralcan_Interconversion_2020} This phenomenon is illustrated by Figure~\ref{fig:DomainGrowth}. We use dashed curves to indicate the prediction for the unrestricted domain growth.

\begin{figure}[b]
    \centering
    \includegraphics[width=\linewidth]{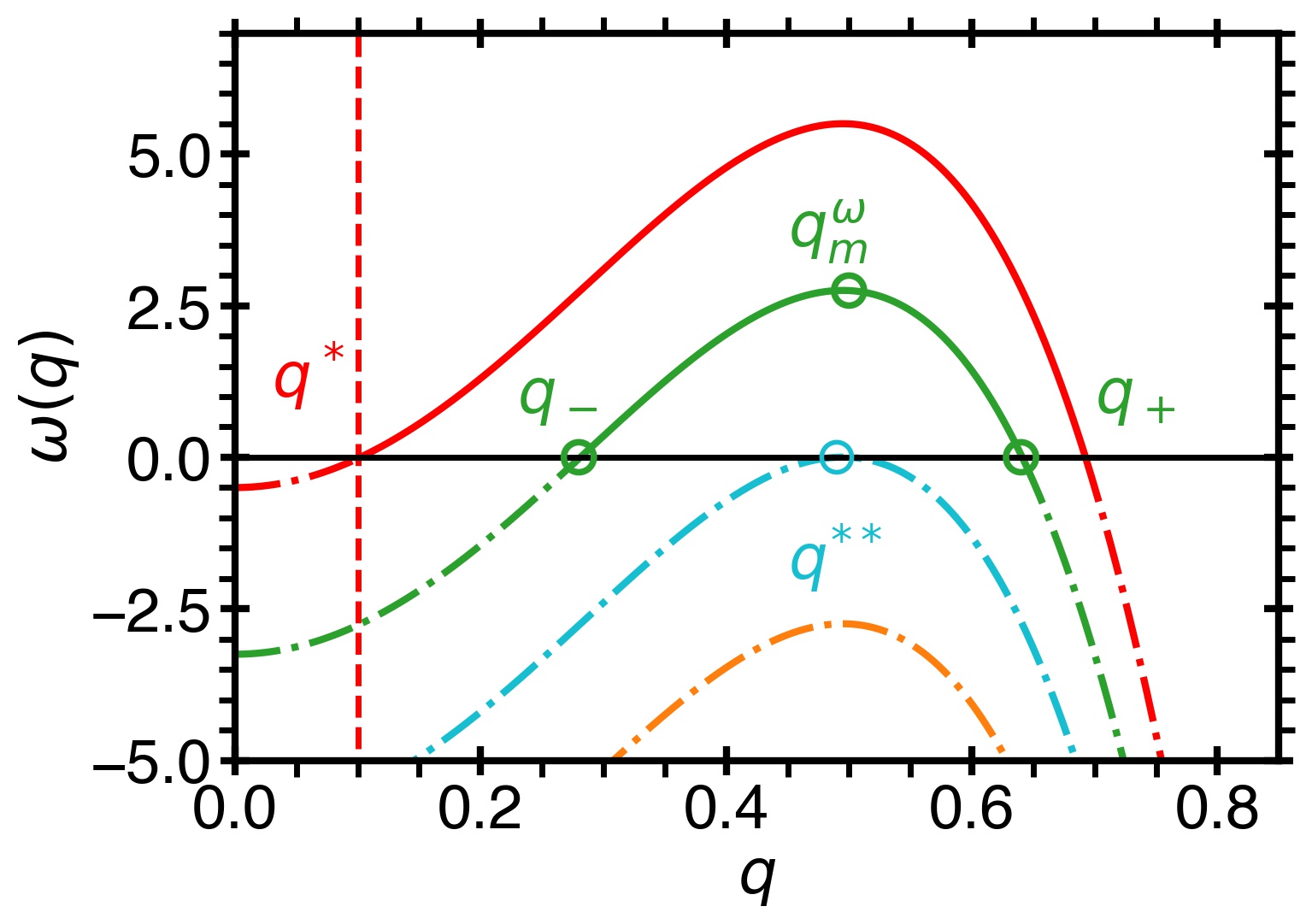}
    \caption{The effect of increasing interconversion force on the phase domain growth rate for $M=100$, $L=1$, and $\Delta\hat{T}=-0.5$. The red dashed line corresponds to the inverse maximum size of the phase domain on the length scale of the simulation box, $q^*$. When $q_- > q^*$ microphase domains will form. Alternatively, when $q_- < q^*$, the size of the simulation box will cut-off the growing phase domains. The conditions where $\omega(q)<0$ (dashed-dot portions of the curves) corresponds to non-growing wavenumbers. As the rate of forceful interconversion increases, the growth rate is shifted down from the onset of phase separation where $q_- = q^*$ (red, $K=1$), to the microphase region (green, $K=3.75$), to the termination point of domain growth (blue, $K=6.5$) where $q_- = q_m^\omega = q_+ = q^{**}$, and to the no growth regime for any wavenumber (orange, $K=9.25$). }
    \label{fig:Growth_Star}
\end{figure}

\begin{figure}[t]
    \centering
    \includegraphics[width=\linewidth]{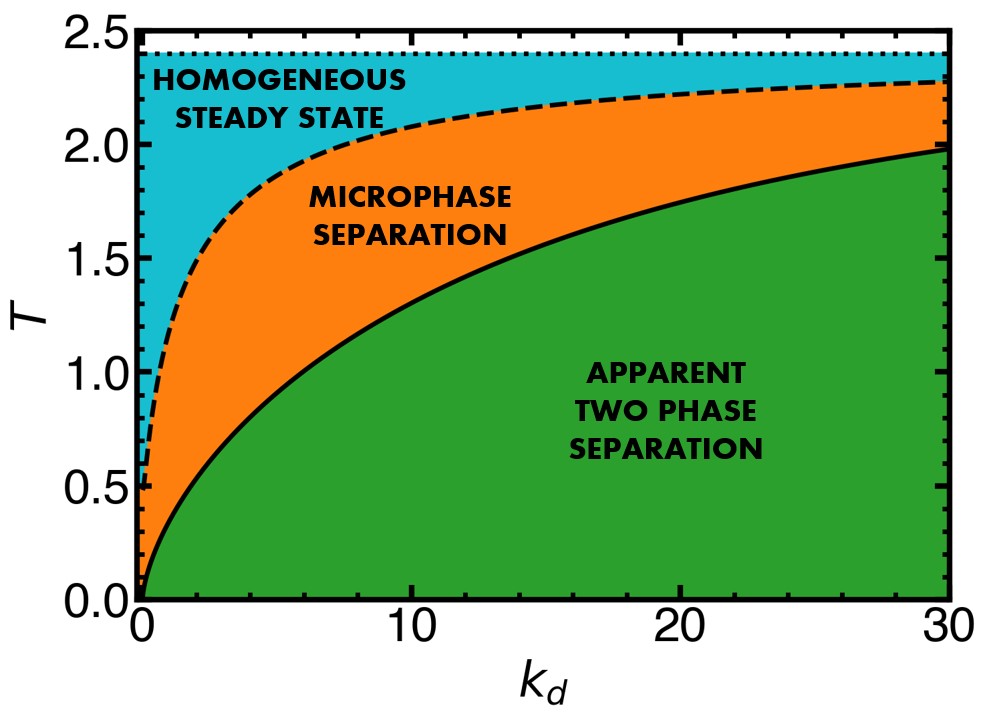}
    \caption{Microphase separation onset temperature $T=T^*$ (solid black) and the termination of microphase separation temperature $T=T^{**}$ (dashed black), Eqs.~(\ref{Eq-OnsetProb} \& \ref{Eqn_DoubleStar}) respectively, for $T_\text{c}=2.4$ and $R_\infty = 1/q^*= 7.1$. At any given value of $k_d$ and $T$, the blue area, above $T^{**}$, indicates a homogeneous steady state region (in which only fluctuational inhomogeneities are present), the orange area, between $T^{*}$ and $T^{**}$, indicates the two phase region where the steady-state size of the phase domain is restricted by the rate of forceful interconversion, and the green area, below $T^*$, indicates the apparent two phase region where the steady-state size of the phase domain is restricted by the size of the system.}
    \label{Fig-Shaded_Region}
\end{figure}

In this case, the line separating the apparent two phase region ($R_\infty > 1/q^*$) and the microphase region ($R_\infty \le 1/q^*$) is found from the first root of the phase domain growth rate - see Figure~\ref{fig:Growth_Star} and Eq. (\ref{Eq-qcmin_first_order}). The line interpreted as the onset of microphase separation (where $R_\infty = 1/q^* \sim  \ell$, $T=T^*$) is determined as
\begin{equation}\label{Eq-OnsetProb}
    \frac{1}{k_d^2} = \frac{2}{\hat{T}^*}\left(\frac{-\Delta\hat{T}^*}{T^*}\right)^2\left[1-\sqrt{1-\frac{(q^*)^2}{(-\Delta\hat{T}^*)}^2}\right]
\end{equation}
Figure~\ref{Fig-Shaded_Region} illustrates the steady-state phase diagram for this system, showing the curve, $T=T^*$, that separates the apparent two phase region from the microphase region. The onset temperature $T^*$ could be viewed as the length-scale dependent critical temperature of microphase separation in the nonequilibrium system. In addition, the time of liquid-liquid phase separation can be quantitatively determined from the characteristic growth rate, $\tau_\text{LLPS}\approx 1/\omega(q^*)$ given by Eq.~(\ref{Eqn_ChiralAmp}) and evaluated at the onset of phase separation when $q_-\approx q^*\sim 1/\ell$. When the temperature is less than the onset temperature $T^*$ liquid-liquid phase separation will take place in the coarsening regime. However, when $T=T^*$, then the time of phase separation will become infinite.

\begin{figure}[t]
    \centering
    \includegraphics[width=\linewidth]{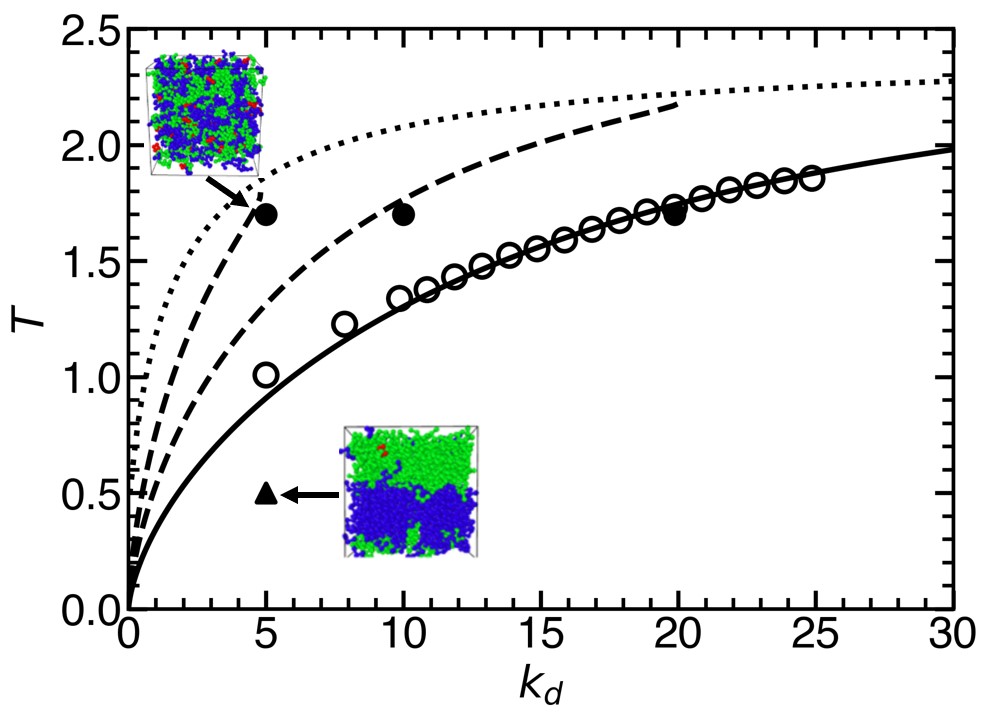}
    \caption{The conditions for microphase separation at a length scale, $R_\infty = 1/q_{-}$. The solid curve depicts the onset temperature of microphase separation, $T=T^*$, at the scale of the computational box, Eq.~(\ref{Eq-OnsetProb}), for $R_\infty = 1/q^* = 7.1$ and $T_\text{c} = 2.4$. The two dashed curves depict the lines of steady-state domain size that are smaller than the size of the computational box, $R_\infty = 3.6$ (lower) and $R_\infty = 2.0$ (upper), while the dotted curve depicts the growth-termination temperature, $T=T^{**}$, Eq.~(\ref{Eqn_DoubleStar}). The symbols are computational data (from the observed size of the phase domain) in the dissipative-force formulation of the chiral model reported by Uralcan \textit{et al.}\cite{Uralcan_Interconversion_2020} for pressure $P = 0.1$. The symbols indicate the onset of phase separation (open circles), the microphase region ($R_\infty = 1/q^*$, closed circles), and the two phase region ($R_\infty > 1/q^*$, triangle). The two simulation snapshots illustrate the behavior of the system composed of A-rich (green), B-rich (blue), and intermediate (red) states of the chiral model. \cite{Uralcan_Interconversion_2020}}
    \label{Fig_TstarScaling}
\end{figure}

In addition to the onset of microphase separation, another phenomenon takes place when the source of forceful interconversion shifts the entire phase domain growth rate below zero. In this scenario, when the growth rate becomes negative for all wavenumbers, a homogeneous steady state exists, in which no phase domain growth will occur. The point at which the maximum of the growth rate crosses zero, $\omega(q_m^\omega)=0$, is characterized through the termination temperature, $T^{**}$. At this point, the three characteristic wavenumbers merge into a single point, $q_m^\omega = q_{-} = q_{+} = q^{**}$ - see Figure~\ref{fig:Growth_Star}. For the dissipative chiral model the termination probability is given in the first order approximation by
\begin{equation}\label{Eqn_DoubleStar}
    \frac{1}{k_d^2} = 2\left(\frac{\Delta \hat{T}^{**}}{\hat{T}^{**}}\right)^2\left[\hat{T}^{**}+\frac{\Delta\hat{T}^{**}}{4}\right]^{-1}
\end{equation}
The termination line, $T= T^{**}$ is also shown in Figure~\ref{Fig-Shaded_Region}. The growth-termination temperature is restricted by the correlation length such that the microphase separation domain is constrained between the size of the computational box and the correlation length of the order-parameter fluctuations. Figure~\ref{Fig_TstarScaling} shows the termination line along with three lines of constant domain size (wavenumber).

\section{Accounting for the Heat and Volume Change of Reaction: Liquid Polyamorphism \label{Sec_Polyamorphism}}
The chiral model with interconversion of enantiomers is one of the simplest examples of liquid polyamorphism, the existence of two liquid states in a single-component substance \cite{Stanley_Liquid_2013,Tanaka_Liquid_2020,Caupin_Minimal_2021}. Chemical-reaction equilibrium constraints the number of thermodynamic degrees of freedom, thus allowing the mixture to be considered as a single-component fluid\cite{anisimov_thermodynamics_2018}. Liquid polyamorphism has been found in multiple substances including hydrogen\cite{Knudson_Direct_2015}, isotopes of helium\cite{Schmitt_Introduction_2015,Vollhardt_Superfluid_1990}, liquid carbon\cite{Glosli_Liquid_1999}, sulphur\cite{Hentry_Liquid_2020}, phosphorous\cite{Katayama_Macroscopic_2004}, and cerium \cite{Cadien_Cerium_2013}. In addition, liquid and vitreous polyamorphism has been supported by simulations of various atomistic models\cite{Sastry_Liquid_2003,Xu_Thermodynamics_2006,Bhat_Vitrification_2007,Xu_Monatomic_2009,Lascaris_Search_2014,Lascaris_Diffusivity_2015}. It has been hypothesized that liquid polyamorphism, via the existence of two alternative molecular or supramolecular structures, may explain the remarkable anomalies of the properties in supercooled water\cite{Holten_Liquid_2012,Stanley_Liquid_2013,Gallo_Tale_2016,Caupin_Thermodynamics_2019,Duska_Water_2020,Tanaka_Liquid_2020,Caupin_Minimal_2021}. The existence of the liquid-liquid transition in supercooled water has been demonstrated by simulations of water-like models\cite{Poole_Phase_1992,Holten_Nature_2013,Holten_Two_2014,Palmer_Metastable_2014,Singh_Two_2016,Debenedetti_One_1998,Gallo_Tale_2016,Gonzalez_Comprehensive_2016,Biddle_Two_2017} and supported by experiment\cite{Mishima_Decompression_1998,Kim_Experimental_2020}. Recently, simulations have confirmed the 3D Ising-model universality class of the liquid-liquid critical behavior in the TIP4P/2005 model of water\cite{Debenedetti_Second_2020}.

\begin{figure}[t]
    \centering
    \includegraphics[width=\linewidth]{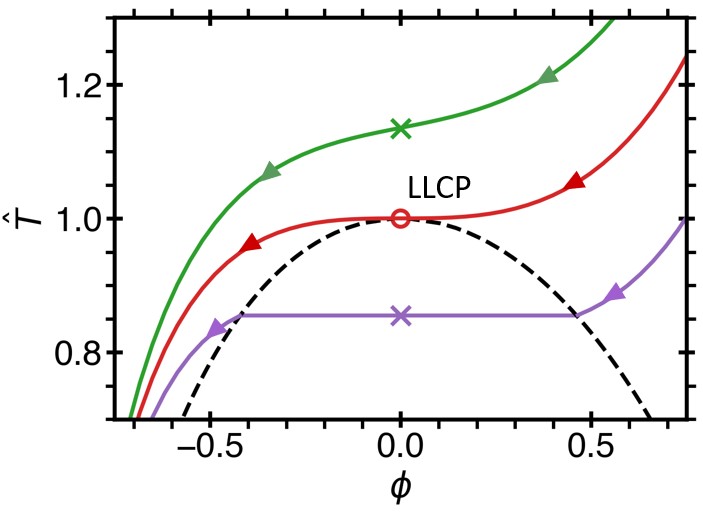}
    \caption{Three hypothesized binary mixture systems exhibiting interconversion of species and liquid-liquid phase separation quenched from high temperature to low temperature (without volume change). The black dashed curve corresponds to the liquid-liquid phase coexistence in this system without interconversion and with interaction energy, $\epsilon = 2$. The open circle indicates the liquid-liquid critical point (LLCP), while the crosses show the locations of $T=T_{BA}$, the points corresponding to 50:50 interconversion for different energy change of reaction. For a system with $T_{BA} = 1.05 T_\text{c}$ (green), no liquid-liquid phase transition will be observed upon quenching. For a system with $T_{BA} = T_\text{c}$ (red), the quenching process passes through the critical point. For a system with $T_{BA}= 0.95T_\text{c}$ (purple), there are two equilibrium solutions for the fraction of interconversion, such that upon quenching to the cross, phase amplification occurs with equal probability of forming an A-rich or B-rich phase.}
    \label{Fig_Int_Quench}
\end{figure}

The interconversion between two alternative molecular or supramolecular states in a single-component substance has been suggested to be a generic cause of liquid polyamorphism\cite{Bertrand_Peculiar_2011,anisimov_thermodynamics_2018,Caupin_Minimal_2021,Singh_Two_2016,Caupin_Thermodynamics_2019,Holten_Two_2014,Biddle_Two_2017,Holten_Nature_2013}. However, in the ``Ising-like'' models (such as the chiral model and the hybrid model considered so far in this work), the Gibbs energy change of the reaction, $\hat{G}_{BA}$, does not depend on pressure or temperature; thus, the forward and reverse equilibrium reaction rates are always equivalent ($\mathcal{K}=1$). Since, in this case, the enthalpy of reaction, given through the Gibbs-Helmholtz relation, $\Delta\hat{H}_{BA} = \hat{T}^2\partial(\ln{\mathcal{K}})/\partial\hat{T}=0$.\cite{LL_Stat_Phys} Therefore, in the one phase region, the equilibrium interconversion between states A and B remains 50:50 being independent of temperature and pressure. However, generally, in most polyamorphic substances, the enthalpy and volume change of the reaction are non-zero, such that, the equilibrium fraction of interconversion depends on temperature and pressure, $\mathcal{K}=\mathcal{K}(T,P)$\cite{anisimov_thermodynamics_2018}. This effect may be incorporated into the time evolution of the order parameter using the complete equation for the reduced chemical potential, Eq.~(\ref{Eqn-ChemPot}). In this case, the solution of Eq.~(\ref{Eqn-GeneralDifferential}) becomes
\begin{equation}
    \phi = \frac{L(\hat{e}+\hat{\upsilon}P-s\hat{T})}{2(K + L\hat{\chi}^{-1}_{q=0})} + \sum_{i} \phi_\infty e^{\omega(q_i)t}\cos(\vec{q}_i\cdot\vec{r})
\end{equation}
where only the infinite time solution of the order parameter is effected by the Gibbs energy of reaction. The phase domain growth rate, $\omega(q)$, remains unaffected and is described by Eq.~(\ref{Eqn-R(q)}). We note that the introduction of the temperature and pressure dependence into the equilibrium interconversion fraction is different from considering the symmetric binary mixture quenched with at a higher concentration of species A (or B). In this case, phase amplification will proceed with a higher probability of forming an A-rich (or B-rich) phase.

\begin{figure}[t]
    \centering
    \includegraphics[width=\linewidth]{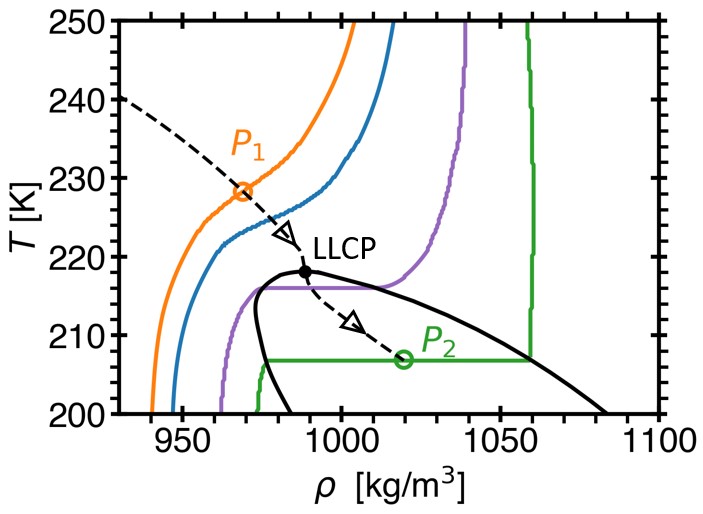}
    \caption{The phase diagram (suggested in ref.\cite{Caupin_Thermodynamics_2019}) for supercooled water that exhibits a liquid-liquid phase transition. A hypothesized quenching process by compression for supercooled water is shown from the one phase region at $P_1=\SI{20}{\mega\pascal}$ (orange) to the two phase region at $P_2=\SI{120}{\mega\pascal}$ (green) along the Widom line (dashed black) which corresponds to a line of constant fraction of interconversion, $\ln{\mathcal{K}}=0$. Two additional isobars are shown for reference at $P = \SI{40}{\mega\pascal}$ (blue) and $P= \SI{80}{\mega\pascal}$ (purple) along with the liquid-liquid coexistence (black). Phase amplification would only be possible in a system where the number of molecules changes to compensate the volume change of the interconversion reaction.}
\label{Fig_WaterPhaseDiagram}
\end{figure}

For the remainder of this section, we will discuss the effects of temperature and pressure on the equilibrium interconversion fraction, $\mathcal{K}$, and how they could affect the possibility of phase amplification. To do so, we will use the hypothesized phase diagram of supercooled water that a exhibits liquid-liquid phase transition, see Figure~\ref{Fig_WaterPhaseDiagram}, to illustrate a real system. We will also make predictions for how phase amplifications could be observed in such a system.

We begin with a discussion of the effect of temperature on the equilibrium fraction, $\mathcal{K}$, in three different hypothesized polyamorphic systems, characterized by three different heat of reactions, $\Delta\hat{H}_{BA}$, in which the two alternative phases have the same density. In this system there are two characteristic energies, $\hat{e}$ and $\epsilon$. We consider the heat of reaction, $\Delta\hat{H}_{BA}=\hat{e}$, which defines the temperature dependence of the equilibrium interconversion fraction, and the heat of mixing $\Delta\hat{H}_{\text{mix}}\propto\epsilon$, given by Eq.~(\ref{Eqn_f0}), which defines the critical temperature of liquid-liquid demixing. Suppose that the $\hat{e} > 0$ such that cooling favors the formation of species B. Depending on the relation between the reference temperature of interconversion $T_{BA}$, at which $\ln\mathcal{K}=0$ (corresponding to 50:50 interconversion), and the critical temperature of demixing one can observe different scenarios upon quenching below $T_\text{c}\propto\epsilon$, as shown in Figure~\ref{Fig_Int_Quench}. In a system with $T_{BA} > T_\text{c}$, the liquid-liquid phase transition will not be observed. If the system has $T_{BA} = T_\text{c}$, the system follows the interconversion fraction to the phase enriched with B without phase coexistence. In a system with $T_{BA} < T_\text{c}$, the system crosses the first-order liquid-liquid phase transition at the point where the interconversion fraction of B is smaller than 50\%. Depending on the final location of the quenching point, phase amplification will occur with a preference to the formation of a single A-rich or B-rich phase. When the final location of the quench is at $\phi = 0$ (corresponding to a 50:50 interconversion rate) as shown in Figure~\ref{Fig_Int_Quench}, phase amplification will occur randomly without a preference to an alternative phase, as discussed in Sec. \ref{Sec_Apps}. However, due to the difference between bulk energies of species A and B, we predict that for one stable phase to grow at the expense of another phase, energy must be supplied or removed from the system. Therefore, either experimentally or computationally, this process must be conducted in a heat reservoir and the thermal conductivity of the system must be faster than the interconversion rate.

Next, we consider a system in which the two alternative species have different molecular volumes and the equilibrium interconversion fraction depends on pressure. In this case, the volume change of the interconversion reaction is not zero. For example, we consider the hypothesized phase diagram of supercooled water suggested by Caupin and Anisimov\cite{Caupin_Thermodynamics_2019} and shown in Figure~\ref{Fig_WaterPhaseDiagram}. It has been suggested the liquid-liquid phase separation in supercooled water is caused by the interconversion between two alternative supramolecular structures\cite{Caupin_Thermodynamics_2019,Holten_Liquid_2012,Duska_Water_2020}. Here, we illustrate a quench by compression from $P_1=\SI{20}{\mega\pascal}$ in the one phase region to $P_2=\SI{120}{\mega\pascal}$ below the critical temperature of demixing along the Widom line for water, where $\ln\mathcal{K}=0$.\cite{Caupin_Thermodynamics_2019,Holten_Two_2014,Singh_Two_2016,Biddle_Two_2017,anisimov_thermodynamics_2018} This condition is preserved along this path because the change in the temperature is compensated by the corresponding change in the pressure as predicted by Eq.~(\ref{Eq_GBA}). In this case, we predict that phase amplification may only occur in the presence of a specific ``barostat'' that supplies or removes molecules to compensate the volume change of the interconversion reaction. Without such a barostat, phase amplification would not be possible. Further elaboration of the developed approach accounting for both the heat and volume change of the interconversion reaction would be desirable.

We note that even though phase amplification may be predicted to occur if there is no volume change of interconversion, there are a variety of factors (like the size of the system, distance from the critical temperature, and the rate of interconversion) that may increase the probability of a metastable interface to form between phases \cite{Shum_Phase_2021}. The effect of possible phase amplification has not been discussed in previous simulations of models for polyamorphic substances, but it could be a factor effecting the possibility of equilibrium phase separation to be observed in both experiment and simulations. 

Since the order parameter in polyamorphic systems is a hybrid containing both conserved and nonconserved components, the coupling between these two components affects the phase domain growth in both nucleation and spinodal decomposition regimes\cite{Takae_Role_2020}. This important feature of the dynamics of liquid polyamorphism could also be elaborated within the framework of interconversion of molecular or supramolecular species.

\section{Effects of Critical Fluctuations of the Order Parameter \label{Sec_CritFlucts}}
As was emphasized in Sec. \ref{Subsec_HybridModel}, the Ising and lattice-gas models are mathematically equivalent. It was later proven that all ﬂuids exhibiting phase separation, whether simple or complex, belong to the same class of critical-point universality in thermodynamics as the Ising
model \cite{fisher_scaling_1983}. Within the same universality class, systems demonstrate the same critical singularities and the same critical equation of state, provided that the appropriately defined order parameter has the same symmetry. The one-component-vector order parameter (the magnetization of the Ising model) and the scalar order parameter (the density of fluids) posses the same $\mathbb{Z}_2$ up-down symmetry \cite{kogut_introduction_1979,senthil_z_2_2000}. 

According to Halperin and Hohenberg\cite{hohenberg_theory_1977}, the Ising and lattice-gas models belong to the different dynamic universality classes, A and H respectively. The relaxation of the conserved order-parameter (density) in the lattice-gas model is controlled by diffusion, while the relaxation of the nonconserved order-parameter (the fraction of spins pointing up or down) in the Ising model is controlled by the flipping of spins (spin ``interconversion''). In addition, the mobility, $M$, in fluids diverges near the critical point as described by the mode-coupling theory\cite{Ohta_ModeCoupling_1975}, while the interconversion kinetic coefficient, $L$, is constant\cite{hohenberg_theory_1977}.

The mean field approximation does not properly incorporate the effects of diverging fluctuations in the critical region. This part of the phase diagram is roughly defined by the region where the correlation length of the order-parameter fluctuations is significantly larger than the distance between molecules (the Ginzburg criterion\cite{LL_Stat_Phys_II}). In practice, the region where physical properties of fluids are significantly modified by fluctuations can roughly be estimated as $\Delta\hat{T} \lesssim 0.1$.\cite{anisimov_critical_1991} Fluctuation-induced effects are described by the renormalization group (RG) theory and the scaling theory of critical phenomena \cite{fisher_scaling_1983,Goldenfeld_Lectures_1992,hohenberg_theory_1977,Wilson_RG_1974}. A comprehensive theory of phase transitions in the presence of molecular interconversion in the vicinity of the critical point has not yet been developed. In this section, we present simple scaling arguments on the behavior of such systems in the approximation of the first-order epsilon expansion of the RG theory in powers of $\epsilon = 4-d$, where $d$ is the system dimensionality. In this approximation, the Landau-Ginzburg free energy functional in the form of Eq.~(\ref{Eqn_LGfunc}), used in the description of spinodal decomposition, corresponds to the Ornstein-Zernike correlation function, in which the susceptibility is proportional to the square of the correlation length, $\hat{\chi}_{q=0} = \xi^2/\kappa$.\cite{LL_Stat_Phys} 

\begin{figure}[t]
    \centering
    \includegraphics[width=\linewidth]{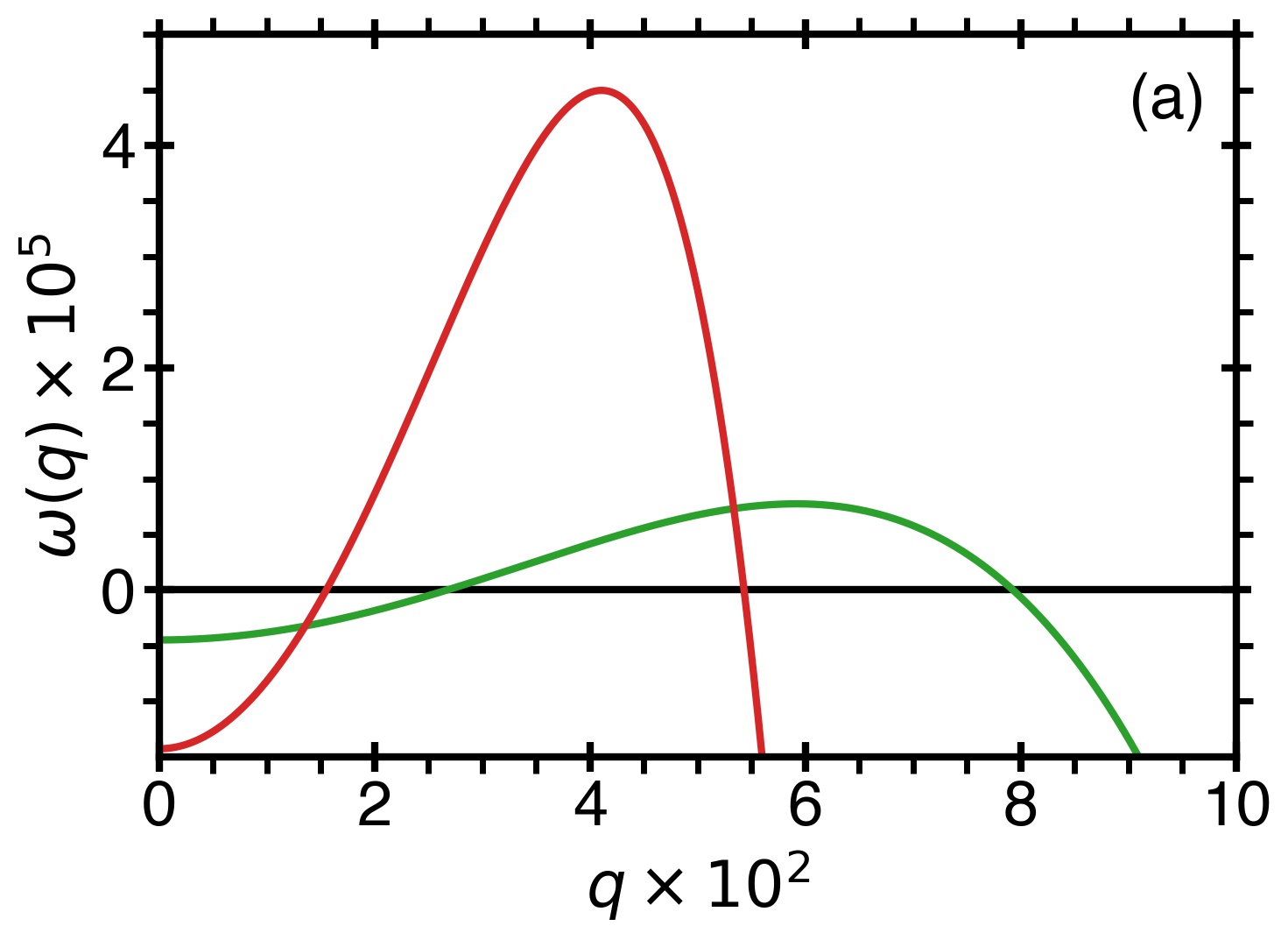}
    \includegraphics[width=\linewidth]{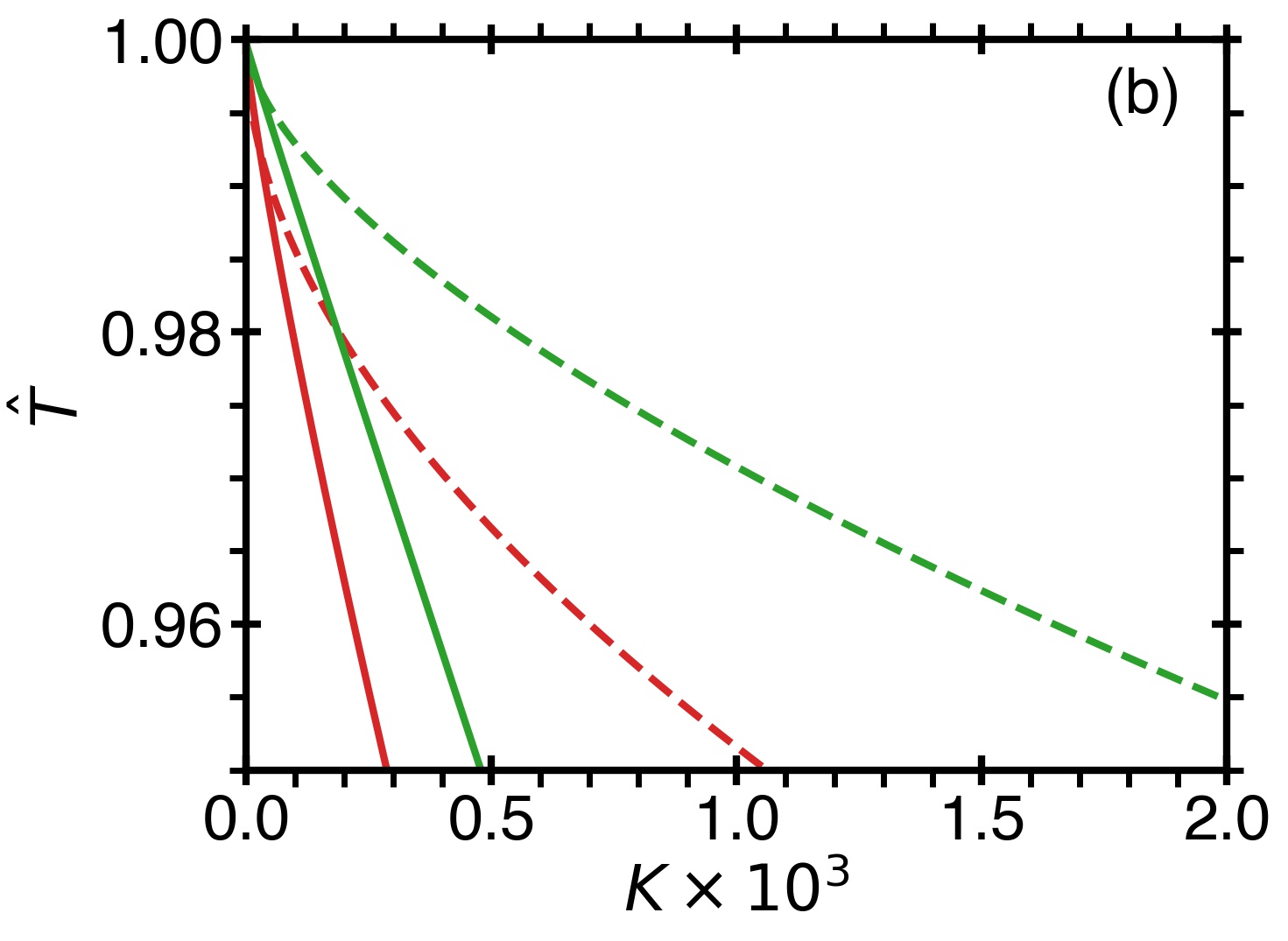}
    \caption{a) Characteristic phase domain growth rate in the vicinity of the critical point ($\Delta \hat{T} = -0.001$) for $M_0 = 1$, $L = 0.002$, and $K = 2.25\times 10^{-5}$ calculated through Eq.~(\ref{Eqn-R(q)_old}), (red curve) with use of the diverging molecular mobility, Eq.~(\ref{DivergingMobility}), and scaling inverse susceptibility in the first order epsilon  expansion, $\hat{\chi}_{q=0}^{-1}\sim |\Delta\hat{T}|^{-\gamma}$ with $\gamma = 1+\epsilon/6$ ($\epsilon=4-d$). The mean field approximation is shown by the green curve, Eq.~(\ref{Eqn-R(q)_old}). b) The onset (red solid curves, Eq.~(\ref{Eqn_Kstar})) and termination (red dashed curves, Eq.~(\ref{Eqn_Kdstar})) of microphase separation affected by critical fluctuations for $M=1$, $L=0.01$, $\ell=100$, $\nu = 1/2+\epsilon/12$. The mean field approximation is shown by the green curves.}
    \label{Fig_AmpScal}
\end{figure}

As the system approaches the critical point, the susceptibility, $\hat{\chi}_{q=0}\sim (\partial^2f_0/\partial\phi^2)^{-1}$, diverges as $|\Delta T|^{-\gamma}$ and the correlation length diverges as $\xi \sim |\Delta\hat{T}|^{-\nu}$, where in the first-order RG epsilon expansion for the 3$d$ Ising-model universality class $\gamma = 1 + \epsilon/6 = 7/6$ and $\nu = \gamma/2 = 1/2 + \epsilon/12 = 7/12$.\cite{Wilson_Fisher_1972} The actual theoretical and most accurate experimental values for the critical exponents differ from those given by the lowest approximation of the RG theory, such that $\gamma = 1.24$ and $\nu = 0.63$.\cite{Anisimov_1974,Agayan_Crossover_2001} The difference between $\gamma$ and $2\nu$ appears only in the second order epsilon expansion, in which $\gamma = \nu(2 - \epsilon^2/54)$.\cite{Wilson_Fisher_1972} However, these differences only marginally change the behavior of the phase domain growth rate presented in Figure~\ref{Fig_AmpScal}a.

There is another effect, the fluctuation induced divergence of the mobility, which is only relevant for the dynamic universality class H. In the vicinity of the critical point, the mode-coupling theory, strongly supported by accurate experimental studies, predicts the divergence of the molecular mobility\cite{Kostko_dynamics_2007}
\begin{equation}\label{DivergingMobility}
    M = M_0\xi^{1-z_\eta}K(q\xi)\left[1+\left(\frac{q\xi}{2}\right)\right]^{z_\eta/2}
\end{equation}
where $M_0 = k_BT/(6\pi\eta_0\kappa)$ is the non-diverging mobility in the mean field approximation; $\eta_0$ is the amplitude of the dynamic viscosity, which weakly diverges as $\eta = \eta_0\xi^{z_\eta}$ ($z_\eta = 1/19$ in the first RG epsilon expansion\cite{Halperin_Divergent_1974,Siggia_dynamics_1976,Sengers_Transport_1985}, and $K(q\xi)\equiv K(x) = [3/(4x^2)](1+x^2+[x^3-x^{-1}]\arctan x)$ is the Kawasaki function\cite{Kawasaki_1976}. The net effect is that the mutual diffusion coefficient asymptotically close to the critical point in the limit $q\to 0$ vanishes as $D_m = M\hat{\chi}^{-1}_{q=0}\sim |\Delta T|^{\nu-z_\eta}$. \cite{Sengers_2006,Onuki_2002} Therefore, the phase domain growth rate given through Eq. (\ref{Eqn-R(q)_old}) incorporates the form of $\hat{\chi}^{-1}_{q=0}$ from scaling theory and $M$ from mode coupling theory, Eq. (\ref{DivergingMobility}).

The comparison between the prediction for the growth rate, Eq.~(\ref{Eqn-R(q)_old}), in the mean field approximation and in scaling theory is shown in Figure~\ref{Fig_AmpScal}a. This figure depicts a significant effect of critical fluctuations on the growth rate. The wavenumber corresponding to the steady-state domain size is shifted toward smaller wavenumbers and the maximum of the growth rate is much stronger, indicating that the domain size growth into steady-state microphase domains will be faster and result in larger domain sizes in the steady-state regime. The structure factor in the scaling regime is given by Eq.~(\ref{Eqn-SQTFinal}) with use of the scaling growth rate and diverging inverse susceptibility, $\hat{\chi}_{q=0}^{-1}\sim |\Delta\hat{T}|^{-\gamma}$ in Eq.~(\ref{Eq-tdepfdp}).

In addition, both the onset and termination of microphase separation are affected by critical fluctuations. In particular, from the growth rate, $\omega(q)$, it can be shown from Eq.~(\ref{Eqn-R(q)_old}) that
\begin{equation}\label{Eqn_Kstar}
    K(\hat{T}^{*}) = L\hat{\chi}^{-1}_{q=0}-(M\hat{\chi}^{-1}_{q=0}+L\kappa)(q^*)^2 - M\kappa(q^*)^4
\end{equation}
where $\hat{T}^*=T^*/T_\text{c}$ is the reduced temperature at the onset. Therefore, the magnitude of forceful interconversion at the onset of microphase separation scales with proximity to the critical point as $K\propto L\hat{\chi}^{-1}_{q=0}\sim |\Delta \hat{T}^*|^{2\nu}$. In contrast, in the mean field approximation, $K\propto |\Delta\hat{T}^*|$. In addition, the magnitude of forceful interconversion at the termination of microphase separation ($\omega(q_m^\omega)=0$) is given by
\begin{equation}\label{Eqn_Kdstar}
    K(\hat{T}^{**}) = \frac{(-M\hat{\chi}^{-1}_{q=0}+L\kappa)^2}{4M\kappa}
\end{equation}
where $\hat{T}^{**}=T^{**}/T_\text{c}$ is the reduced temperature at the termination point, and thus, $K\propto D_m\hat{\chi}^{-1}_{q=0}\sim |\Delta\hat{T}^{**}|^{3\nu}$, while in the mean field approximation, $K\propto |\Delta\hat{T}^{**}|^{2}$. As shown in Figure~\ref{Fig_AmpScal}b, the effect of critical fluctuations lowers the amplitude of the onset and termination lines. Away from the critical point, these lines should converge with their mean field predictions. Lastly, we note that for the dissipative force formulation of the chiral model unlike the strong effect of the critical fluctuations on the phase domain growth rate, the shape of the $T^*(k_d)$ curve is only marginally affected by accounting for critical fluctuations.

\section{Conclusions}
We have presented a phenomenology of phase transitions affected by both natural and forceful molecular interconversion. The model is applicable to systems where the order parameter possesses both conserved and non-conserved dynamics, such as a hybrid of the incompressible binary-liquid lattice model and the Ising model for an anisotropic ferromagnet. We have shown that a source of forceful interconversion may cause microphase separation, while in the absence of forceful interconversion, the competition between diffusion and natural interconversion dynamics results in the phenomenon of phase amplification, when one phase grows at the expense of another stable phase.

The model has promising applications to more complex systems with interconversion of molecules or supramolecular structures. For example, by incorporating the pressure and temperature dependent fraction of interconversion, the developed approach could be applied to the description of systems exhibiting liquid and vitreous polyamorphism (Sec. \ref{Sec_Polyamorphism}). In addition, the phenomenon of microphase separation could be used to clarify the formation of membraneless organelles\cite{Shakhnovich_Organelles_2019,Dar_Phase_2020}. Just like molecular interconversion in the presence of an external energy source, the interconversion of proteins from a folded to an unfolded state in biological systems, as well as liquid-liquid phase separation induced by polymerization and by the interconversion of supramolecular structures,\cite{Shumovskyi_Modeling_2021} could be also considered using the developed approach.

Future work could be to understand the role of the surface energy in both phase amplification and microphase separation phenomena. In addition, a more in depth study of the morphology of microphase separation could lead to an understanding of the nonequilibrium microemulsion structures formed due to the presence of the source of forceful interconversion.

Lastly, microphase structures caused by forceful interconversion can be considered as an example of dissipative structures. It would be promising to connect the developed approach with the general theory of dissipative structures of Prigogine \textit{et. al.} \cite{nicolis_self-organization_1977} It could also have cross-disciplinary applications to other nonlinear phenomena like hydrodynamic instabilities \cite{wang_progress_2019}, bifurcations, and catastrophe theory\cite{Arnold_Catastrophe_1984}. 

\acknowledgements{We thank Sergey Buldyrev, Fr\'ed\'eric Caupin, Pablo Debenedetti, Nikolay Shumovskyi, and Bet\"ul Uralcan for helpful discussions and encouragements. This work is a part of the research collaboration between the University of Maryland, Princeton University, Boston University, and Arizona State University supported by the National Science Foundation. The research at the University of Maryland was supported by NSF award no. 1856479.}

\section*{Data Availability}
Data sharing is not applicable to this article as no unpublished data were created or analyzed in this study.

\section*{References}
\bibliography{ref}

\end{document}